%% file: main.tex
\documentclass[nonacm,manuscript]{acmart}

\input{0_macro}

\usepackage{enumitem}
\usepackage{soul}
\usepackage{listings}
\usepackage{ulem}

\AtBeginDocument{%
  }

\begin{document}

\title[Contextual Gists for Blind and Low Vision Screen Reader Users’ Understanding of Dynamic User Interfaces]{\systemname{}: Contextual Gists for Blind and Low Vision Screen Reader Users’ Understanding of Dynamic Interfaces}


\author{Ritesh Kanchi}
\email{rkanchi@g.harvard.edu}
\orcid{0009-0006-7978-0821}
\affiliation{
    \department{Engineering and Applied Sciences}
    \institution{Harvard University}
    \city{Cambridge}
    \state{Massachusetts}
    \country{USA}
}
\author{Jianna So}
\email{jiannaso@g.harvard.edu}
\orcid{}
\affiliation{
    \department{Engineering and Applied Sciences}
    \institution{Harvard University}
    \city{Cambridge}
    \state{Massachusetts}
    \country{USA}
}
\author{Krzysztof Z. Gajos}
\email{kgajos@g.harvard.edu}
\orcid{0000-0002-1897-9048}
\affiliation{
    \department{Engineering and Applied Sciences}
    \institution{Harvard University}
    \city{Cambridge}
    \state{Massachusetts}
    \country{USA}
}


\begin{abstract}
\input{abstract}
\end{abstract}


\begin{teaserfigure}
    \centering
  \includegraphics[width=\textwidth,alt={A simplified diagram of the Coral system. On the left, a pink illustration of a shopping webpage represents the User Interface and Screen Reader. An arrow leads into Coral's Interface Context, shown as three vertically stacked components: Structural Context, represented by a yellow block of webpage code; Semantic Context, represented by a green panel listing the recognized patterns Product Grid, Faceted Search, Header Navigation, Sidebar Filtering, and Call to Action; and Visual Context, represented by an orange rendering of the webpage. Coral combines this information with User Context, comprising an Experience Store and Interface and Interaction History. The Experience Store is depicted as a blue cloud containing previously encountered interface categories, including E-Commerce, Social Media, and Business. Interface and Interaction History is depicted as overlapping snapshots of the shopping webpage. A downward arrow leads to a purple Importance Evaluation panel containing State Comparison, Text Delta, Noise Filtering, and LLM Evaluation. If the information is deemed important, Coral performs Gist Generation and produces the following Screen Reader Output: Store products page features a top navigation bar, a filters sidebar, and a main content area displaying a grid of items like the 15-inch laptop and the 13-inch laptop.}]{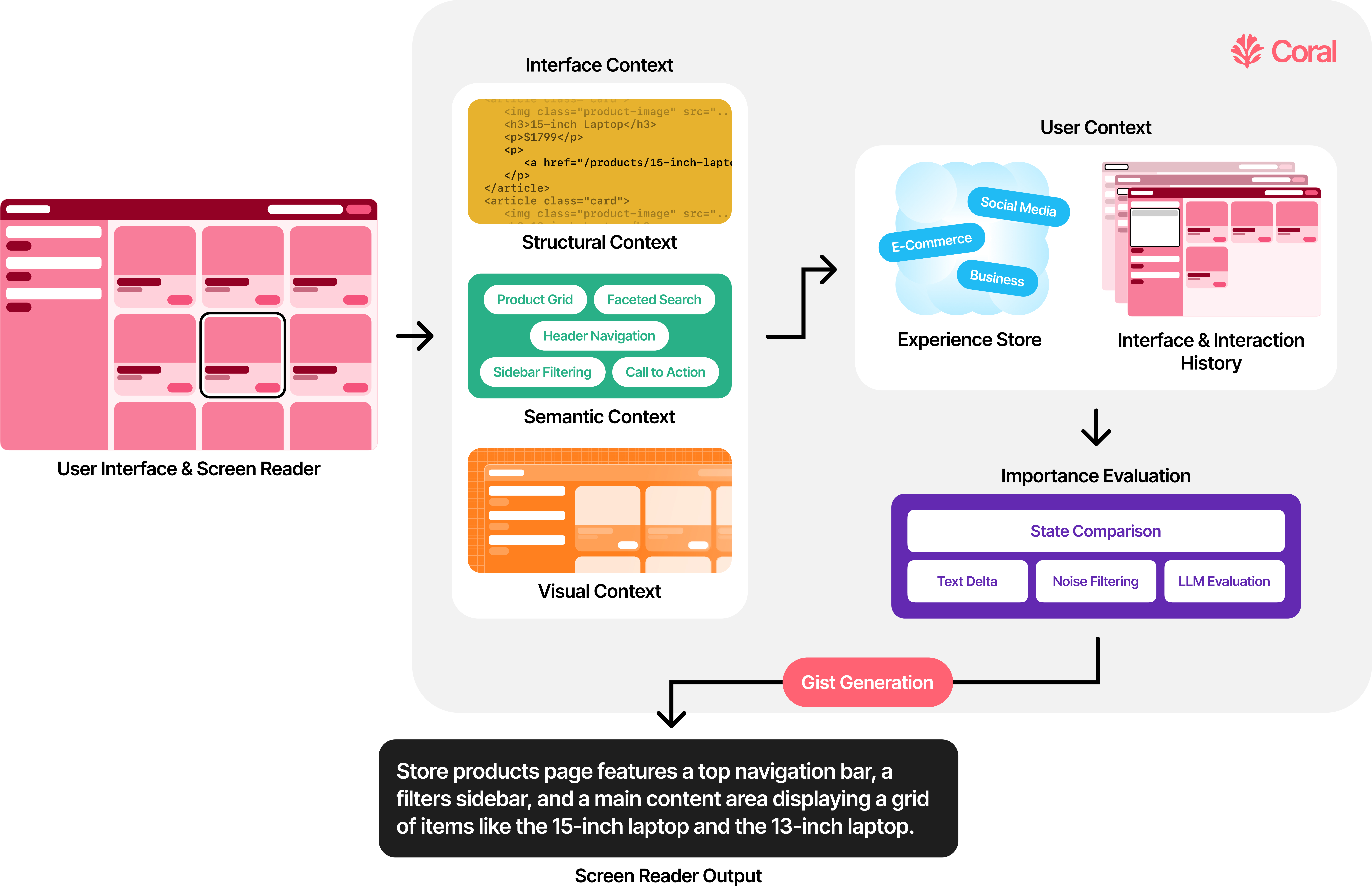}
  \caption{
  \systemname{} generates gists of relevant interface states and changes by synthesizing \textit{interface context}---structural, semantic, and visual evidence about the UI---with \textit{user context} comprising recent interactions, prior interface states and representations of previously encountered interfaces. If the synthesized context is deemed important, \systemname{} narrates the gist through the user's screen reader output.
  }
  \Description{A simplified diagram of the Coral system. On the left, a pink illustration of a shopping webpage represents the User Interface and Screen Reader. An arrow leads into Coral's Interface Context, shown as three vertically stacked components: Structural Context, represented by a yellow block of webpage code; Semantic Context, represented by a green panel listing the recognized patterns Product Grid, Faceted Search, Header Navigation, Sidebar Filtering, and Call to Action; and Visual Context, represented by an orange rendering of the webpage. Coral combines this information with User Context, comprising an Experience Store and Interface and Interaction History. The Experience Store is depicted as a blue cloud containing previously encountered interface categories, including E-Commerce, Social Media, and Business. Interface and Interaction History is depicted as overlapping snapshots of the shopping webpage. A downward arrow leads to a purple Importance Evaluation panel containing State Comparison, Text Delta, Noise Filtering, and LLM Evaluation. If the information is deemed important, Coral performs Gist Generation and produces the following Screen Reader Output: Store products page features a top navigation bar, a filters sidebar, and a main content area displaying a grid of items like the 15-inch laptop and the 13-inch laptop.}
  \label{fig:teaser}
\end{teaserfigure}
%



\maketitle

\input{1_introduction}
\input{2_new_related_work}
\input{3_positionality}
\input{4_interview}
\input{5_design}
\input{5_coral}
\input{6_evaluation}

\input{7_new_discussion}
\input{8_limitations}
\input{9_conclusion}


\bibliographystyle{ACM-Reference-Format}
\bibliography{base,kzg}

\appendix
\input{appendix}

\end{document}

%% file: 0_macro.tex
\usepackage{cleveref}
\usepackage{ifthen}
\usepackage{subcaption}
\usepackage{booktabs}
\usepackage{xspace}
\usepackage{tabularx}
\usepackage{colortbl}
\usepackage{tcolorbox}

\usepackage{listings}
\newcommand{\ie}{\textit{i.e. }}
\newcommand{\eg}{\textit{e.g.,}}

\newcommand{\systemname}{\textsc{Coral}}

\usepackage{xcolor}
\usepackage{tikz}
\usepackage{tabularx}

\definecolor{coralcolor}{HTML}{FF6272}
\definecolor{actioncolor}{HTML}{555555}

\newcommand{\gistpill}[2]{%
  \tikz[baseline=(pill.base)]{
    \node[
      fill=#1,
      text=white,
      rounded corners=5pt,
      inner xsep=5pt,
      inner ysep=3pt,
      font=\sffamily\bfseries\scriptsize
    ] (pill) {#2};
  }%
}

\newcommand{\gisttable}[5]{%
  \par
  \noindent
  \begin{minipage}{\linewidth}
    \vspace*{1\baselineskip}

    \noindent\textbf{Gist #1: #2}
    \par\smallskip
    \noindent
    \begin{tabularx}{\linewidth}{@{}l@{\hspace{0.6em}}X@{}}
      \gistpill{actioncolor}{#3}
        & #4 \\[0.5em]
      \gistpill{coralcolor}{\systemname~Gist}
        & #5
    \end{tabularx}

    \par\vspace*{1\baselineskip}
  \end{minipage}
  \par
}



%% file: abstract.tex
Blind and low vision (BLV) screen reader users construct mental models of user interfaces (UIs) through incremental screen reader interaction, a time-consuming and cognitively demanding process complicated by modern interfaces that may not convey dynamic content accessibly.
We interviewed eleven BLV screen reader users about UI understanding and derived design objectives that informed \systemname{}, a context-aware browser extension. \systemname{} synthesizes interface and user context to generate screen reader-narrated gists
which holistically notify users of interface states and changes relevant to their likely goals and ongoing interaction.
In a comparative evaluation with eight BLV screen reader users, participants used \systemname{} to form initial expectations and interpret interface changes, and reported spending less time and effort manually verifying interaction outcomes.
Together, our findings provide deeper insight into how BLV screen reader users navigate gaps in their UI understanding, and how intelligent support that combines interface and user context can bridge these gaps.

%% file: 1_introduction.tex
\section{Introduction}
\begin{quote}
    \textit{``[\systemname{}] feels like having [a] helper looking over my shoulder and telling me about things I might need to know about [on] some of the crazy [interfaces] out there''} - P7
\end{quote}

User interfaces (UIs) are interactive, stateful, dynamic, spatial, and inherently nonlinear~\cite{giudice_navigating_2018, potluri_examining_2021, chheda-kothary_it_2025, chheda-kothary_understanding_2023, borodin_more_2010, yu_cluttered_2025}. While sighted users can often perceive UI elements, layout, and changes at a glance, blind and low vision (BLV) users primarily access interface content by navigating serially through a screen reader~\cite{design_of_everyday_norman_2013, vision_marr_2010,  giudice_navigating_2018, soper-the-nature-2016}. 
As users encounter and interact with UIs, they construct a \textit{mental model of the UI}: a partial, evolving representation of its structure, state, functionality, and behavior~\cite{mental_models_hci_carroll_1988, design_of_everyday_norman_2013, human_computer_interaction_preece_1994, norman-observations-1987}. 
Shaped by prior knowledge, available interface information, ongoing interactions, and users' goals, mental models support \textit{expectations}: situated predictions about what interface information should be present, what interactions are possible, and how the interface should behave~\cite{design_of_everyday_norman_2013, mental_models_hci_carroll_1988, soper-the-nature-2016, norman-observations-1987, doi-effects-asymmetry-2021, staggers-mental-models-1993}. 
These expectations guide exploration and interaction and provide a basis for evaluating whether interface outcomes align with users' current understanding~\cite{design_of_everyday_norman_2013, doi-effects-asymmetry-2021, soper-the-nature-2016}.

Because screen reader access is incremental, \textit{UI understanding}---constructing, maintaining, adapting, and using mental models of UI---can be a time-consuming, labor-intensive, and cognitively demanding~\cite{mental_models_laird-2010, vision_marr_2010, potluri_examining_2021, bigham-2017-dontknow, borodin_more_2010}.
Screen reader mediation can also make expectations difficult to form and evaluate when necessary evidence is inaccessible, incomplete, or poorly surfaced; users may not know whether missing information is absent or merely inaccessible~\cite{surfers2005, bigham-2017-dontknow, sahib_info_seeking_2012}. 
\textit{Dynamic interface behavior} exacerbates this ambiguity as content is added, removed, or updated during use. Changes may be \textit{interaction-driven} (\eg{} a sidebar opening after a button is clicked, navigating to a page with new UI elements) or \textit{autonomous}, in which the interface changes independently  (\eg{} an alert appearing without a preceding user action)~\cite{bostic2019exploringintersectionswebscience,borodin_more_2010,yu_cluttered_2025}.
When screen readers do not communicate these changes coherently---often because accessibility support is lacking, incomplete, or misused~\cite{bi_accessibility_2022, zhong_screenaudit_2025, martin-large-scale-2024, webaim_million, fok_large_scale_2022}---users may need to re-explore the interface to determine what changed, whether an interaction succeeded, or what the current state is~\cite{borodin_more_2010, bostic2019exploringintersectionswebscience, carvalho-accessibility-usability-problems-2018, bigham-2017-dontknow}. 
Inaccessible content and unobserved changes can leave users' mental model misaligned with the interface state, producing expectation mismatches that require further exploration and adaptation~\cite{norman-observations-1987, liu-considering-2010, design_of_everyday_norman_2013, staggers-mental-models-1993}.

To investigate how BLV screen reader users understand modern UIs, we conducted semi-structured interviews with 11 BLV screen reader users. 
Participants constructed and used mental models from the evidence available through the screen-reader-mediated interface.
On familiar interfaces, they relied on interface-specific mental models; on unfamiliar interfaces, they drew on generalized schemata and recognizable design patterns to form expectations and guide exploration~\cite{human_computer_interaction_preece_1994, design_of_everyday_norman_2013, saei_mental_2010, mental_models_hci_carroll_1988, van-welie-interaction-patterns-ui-2000, dearden-pattern-lang-hci-2006}. Although preceding actions provided causal anchors for interaction-driven changes, participants often manually verified when expected outcomes occurred. In contrast, autonomous changes could remain unnoticed until participants re-explored the interface or received explicit notice. When screen reader output was insufficient, participants sought complementary context from community members, AI tools, and other sources, particularly to understand visual content and spatial relationships.
These findings motivate support that provides contextual evidence for UI understanding beyond access to individual elements. Although prior intelligent systems have enriched interface descriptions or improved accessibility, less work has connected dynamic interfaces to users' expectations, interactions, and prior interface experiences~\cite{peng-diffscriber-2022, surfers2005, yu_cluttered_2025, gubbi_mohanbabu_context-aware_2024, chen-struggle-2026}.

Our interview findings, related work, and prototyping with BLV consultants informed five design objectives instantiated in \systemname{}, a context-aware browser extension that generates gists of relevant interface states and changes through users' screen readers.
To generate these gists, \systemname{}  uses large language models (LLMs) and vision-language models (VLMs) to synthesize \textit{interface context}---structural, semantic, and visual evidence about the UI---with \textit{user context} comprising recent interactions, prior interface states, and representations of previously encountered interfaces. This synthesis allows \systemname{} to describe what changed and where, relate changes to users' actions when supported by available evidence, and draw upon familiar design patterns. Rather than representing users' mental models directly, user context helps \systemname{} infer relationships relevant to their likely goals and ongoing interactions.

We then conducted a comparative within-subjects evaluation with 8 BLV screen reader users performing mock e-commerce tasks (\eg{} finding products, adding products to a cart, checking out), using \systemname{} as a research probe into the design objectives. Participants used its interface-specific information alongside their own schemata to form initial expectations and organize their understanding of unfamiliar interfaces. They reported spending less time and effort manually verifying outcomes when \systemname{} related changes to preceding actions, while comparisons with prior states helped distinguish updated content from stable regions.
Progress feedback signaled that a change occurred before the gist interpreted it, and several participants developed a \textit{Catch-and-Replay} interaction to defer interpretation until they were ready. However, incomplete or insufficiently grounded gists sometimes left expectations unresolved, while \systemname{}'s proactive and asynchronous narration introduced new interpretative, attentional, and coordination demands.
 

In summary, we contribute: (1) an empirical investigation of how BLV users construct and adapt mental models of modern UIs; (2) five design objectives for intelligent support of BLV UI understanding and \textbf{\systemname{}}, a context-aware browser extension that instantiates them through contextual gists; and (3) an evaluation of \systemname{} as a research probe for supporting BLV UI understanding.


%% file: 2_new_related_work.tex
\section{Related Work}
\subsection{Mental Models and BLV UI Understanding}
\label{rw:mental-models}


Mental models have been conceptualized both as internal representations used in reasoning and, within HCI, as users' evolving understandings of interactive systems~\cite{mental_models_laird-1980, mental_models_laird-2010, nature_explanation_craik_1943, norman-observations-1987, design_of_everyday_norman_2013}. We adopt the latter account, treating a user's mental model as a partial and simplified representation shaped by their goals, prior experience, and the information available to them~\cite{mental_models_hci_carroll_1988, design_of_everyday_norman_2013, human_computer_interaction_preece_1994}. Mental models can support effective system use partly by providing a basis for users to form \textit{expectations} about what information should be present, what interactions are possible, and how the system should behave~\cite{design_of_everyday_norman_2013,  mental_models_hci_carroll_1988, liu-considering-2010, soper-the-nature-2016}. Throughout interaction, users compare their experience using the system against these expectations, relying on matches and mismatches to interpret outcomes and, when necessary, revise specific expectations or adapt their broader mental model~\cite{norman-observations-1987, liu-considering-2010, soper-the-nature-2016, mental_models_hci_carroll_1988, human_computer_interaction_preece_1994}. This comparison is central to the ``gulf of evaluation'': after acting, users must perceive, interpret, and evaluate system state to determine whether an outcome successfully aligns with their expectations~\cite{design_of_everyday_norman_2013}. Mental models also support knowledge transfer through \textit{schemata}---generalized knowledge structures, formed through prior experiences with other systems, that guide understanding in unfamiliar environments~\cite{human_computer_interaction_preece_1994, design_of_everyday_norman_2013, saei_mental_2010, mental_models_hci_carroll_1988}. When a new system follows recognizable patterns, users can apply relevant schemata rather than construct an understanding from scratch. However, this transfer can hinder interaction when schema-based expectations do not correspond (or appear to correspond) to the encountered system~\cite{liu-considering-2010, design_of_everyday_norman_2013, van-welie-interaction-patterns-ui-2000, dearden-pattern-lang-hci-2006}.

This tension is especially consequential for UIs, where familiar design patterns support efficient inference not only about what elements are present, but also about how they are organized, how they behave, and how they respond to user actions~\cite{van-welie-interaction-patterns-ui-2000, dearden-pattern-lang-hci-2006}. Because UIs are interactive, stateful, dynamic, and spatial, users must reconcile these expectations with interface-specific evidence to construct mental models of the interface's structure, state, functionality, and behavior~\cite{giudice_navigating_2018, potluri_examining_2021, chheda-kothary_it_2025, chheda-kothary_understanding_2023, borodin_more_2010, yu_cluttered_2025}. 
Sighted users can often draw on visuospatial cues (\eg{} grouping, proximity) to perceive multiple interface elements, regions, and relationships concurrently~\cite{mental_models_laird-1980, vision_marr_2010, design_of_everyday_norman_2013, giudice_navigating_2018}. By contrast, screen reader interaction typically presents interface information serially, requiring BLV users to integrate incrementally encountered elements into mental models that are constructed and adapted over time~\cite{mental_models_laird-2010, vision_marr_2010, potluri_examining_2021}. 
When screen reader-mediated access omits, delays, or poorly communicates relevant information, users may lack the evidence needed to evaluate the resulting interface state with their expectations. The gulf of evaluation may therefore widen, leaving users uncertain whether an interaction succeeded, an expected change occurred, or information is absent rather than inaccessible~\cite{bigham-2017-dontknow, borodin_more_2010}.

Whether users can bridge the gulf of evaluation depends partly on the representation through which the system is available to them. \citet{zhang-interaction-proxies-2017} term this representation the \textit{manifest interface}----the interface directly perceived and manipulated by a person---which differs from the interface intended by its designers and developers.
 For people with disabilities, the manifest interface may differ also vary across modes of access. 
Within \citet{design_of_everyday_norman_2013}'s account, it mediates how the target system and designers' conceptual model become available to users, shaping the mental models they can construct~\cite{doi-effects-asymmetry-2021, norman-observations-1987, design_of_everyday_norman_2013}. Correspondence between a user's mental model and the system depends partly on whether the manifest interface coherently surfaces the information needed for understanding. 
 For BLV screen reader users, the screen-reader-mediated manifest interface may omit, delay, or reorganize information available in the visually rendered interface~\cite{bigham-2017-dontknow, borodin_more_2010}.

Together, this literature establishes that BLV UI understanding depends on constructing mental models through incremental navigation, forming and evaluating expectations about the interface, and interpreting the current interface state. Furthermore, screen reader-mediation can expand the gap between the manifest and intended interfaces---especially when interface state can be ambiguous for BLV users to properly evaluate expectations about the interface.


\subsection{The State and Accessibility of Modern UIs}
Despite established accessibility guidelines, web accessibility barriers remain widespread. The 2026 WebAIM Million report detected Web Content Accessibility Guidelines (WCAG) 2\footnote{\url{https://w3.org/WAI/standards-guidelines/wcag/}} violations on 95.9\% of the home pages among the top million websites, an unexpected increase from 94.8\% after several years of improvement~\cite{webaim_million}.
These violations are compounded by modern web development practices, such as the use of component-based frameworks (\eg{} React\footnote{\url{https://react.dev/}}, Vue\footnote{\url{https://vuejs.org/}}) that rely on client-side rendering and virtual Document Object Model (DOM) architectures which introduce delayed content rendering, dynamically inserted controls, and focus instability~\cite{react-accessibility}. Such dynamic interface behavior may be interaction-driven (\eg{} a form updating after a user clicks submit) or autonomous, where the interface changes independently without a preceding user action(\eg{} an interface adding an alert to the top of the interface).
While the Accessible Rich Internet Applications (ARIA)\footnote{\url{https://www.w3.org/WAI/standards-guidelines/aria/}} suite was designed to improve access to dynamic interface content, increased ARIA usage has not necessarily translated into greater accessibility. While \citet{webaim_million} reported that ARIA usage increased substantially (up 27\% in 2026), they further correlated greater ARIA usage with higher detected errors, suggesting that dynamic interface accessibility depends not only on the presence of such support, but on whether it is implemented properly by developers~\cite{bi_accessibility_2022, zhong_screenaudit_2025, martin-large-scale-2024, webaim_million, fok_large_scale_2022}.

For screen reader users, inaccessible or poorly surfaced dynamic behavior creates ambiguity in perceiving and interpreting interface state, which was already identified to be a challenge within static interfaces~\cite{bostic2019exploringintersectionswebscience, bigham-2017-dontknow, yu_cluttered_2025}. Screen reader users will often navigate with these limitations in mind, expecting interfaces to only be partially interpretable, leading to additional time and effort spent accomplishing tasks~\cite{borodin_more_2010}. These challenges are particularly common in domains such as e-commerce, where complex layouts, inconsistent structure, and retrieval-heavy tasks make navigation and information seeking difficult even when interfaces follow accessibility guidelines~\cite{yu_cluttered_2025, bi_accessibility_2022, carvalho-accessibility-usability-problems-2018}. 

For BLV screen reader users, inaccessible, dynamic, and inconsistently communicated interface states can make it difficult to form and evaluate expectations and maintain aligned mental models. Although accessibility mechanisms such as ARIA can expose dynamic changes, their presence does not guarantee that those changes will be coherently communicated through a screen reader. These limitations motivate complementary support for perceiving and interpreting interface state when conventional accessibility is missing or absent.

\subsection{Intelligent Systems for BLV UI Understanding}
As BLV screen reader users encounter inaccessible interfaces on an everyday basis, many rely upon allies (\eg{} sighted friends and family) and remote visual interpreting services (\eg{} Aira\footnote{\url{https://aira.io/}}) to help them interact and gain more context. More recently, BLV users have turned toward AI tools for this support, primarily due to privacy, independence, and accuracy concerns of humans~\cite{adnin_i_2024, penuela-investigating-2024, hayes-cooperative-privacy-2019, feng-understanding-inform-2024, sharma_before_2025}. This support extends both to general-purpose AI tools (\eg{} ChatGPT\footnote{\url{https://chatgpt.com/}}, Gemini\footnote{\url{https://gemini.google.com}}) as well as bespoke accessibility AI tools (\eg{} Be My AI\footnote{\url{https://bemyeyes.com/bme-ai/}}, JAWS Picture Smart AI\footnote{\url{https://blog.freedomscientific.com/picture-smart-ai-how-we-got-here/}})~\cite{gonzalez_penuela_towards_2025, perera_sky_2025}. 

Researchers have long developed systems that synthesize content beyond individual elements to help BLV screen reader users understand and interact with UIs. Early work by \citet{surfers2005} used web page content to generate gist summaries,  reducing the cognitive effort required for interface-based decision-making and exhaustive exploration. \citet{peng-diffscriber-2022} developed \textit{Diffscriber}, which describes visual presentation changes made by collaborators, allowing BLV authors to review changes without re-exploring the presentation to determine what changed. 
Interface context can support UI understanding at different scopes: within an interface state, by summarizing content into a concise gist, and across states, by identifying what changed. In both cases, contextualizing element-level information reduces the need for users to reconstruct understanding through exhaustive exploration.
Recent LLM- and VLM-based systems have extended the use of interface context by synthesizing interface information at different scopes. \citet{yu_cluttered_2025} used an LLM to interpret the broader semantics and structure of e-commerce pages and reorganize or regenerate their HTML, producing a cleaner navigation hierarchy for screen reader use. 
At the element level, web images often lack sufficient alternative text~\cite{webaim_million, gubbi_mohanbabu_context-aware_2024, martin-large-scale-2024, gaggi-accessibility-vi-2019, carvalho-accessibility-usability-problems-2018}. \citet{gubbi_mohanbabu_context-aware_2024} incorporated surrounding page content into a VLM for image description generation, finding that BLV users preferred context-aware descriptions for their greater quality, relevance, and plausibility. 
Together, these systems demonstrate how broader interface context can make both individual content and overall interface structure more interpretable through screen readers.

However, the context necessary for shaping UI understanding extends beyond the interface itself to the user: their goals, ongoing interactions, and prior experiences. Prior work has demonstrated how interaction histories can represent aspects of this user context.
\citet{wexelblat_footprints_1999} argued that web page transitions---made as users travel from page to page---can externalize information relationships users perceive across pages.
Similarly, \citet{gajos07:automatically} used traces of prior interactions to adapt interfaces, while subsequent work on \textit{Personalized Dynamic Accessibility} argued that such adaptations should reflect users' abilities and evolving environments~\cite{gajos_personalized_2012}. For BLV users, \citet{potluri_examining_2021} suggested comparing unfamiliar interfaces to frequently used interfaces could enable richer semantic descriptions for UI understanding. 
More recent AI agents have incorporated immediate aspects of user context into assistance. \citet{peng-morae-2025} introduced \textit{Morae}, which proactively-pauses to elicit BLV users' preferences during UI interaction, supporting preference expression,decision-making, and task completion compared to general-purpose agents. Similarly, \citet{chen-struggle-2026} developed \textit{AskEase}, an on-demand assistant that combines interface and user context to generate guidance for screen reader users. 

Prior research shows that intelligent systems can support BLV UI understanding by contextualizing individual elements within the broader interface, synthesizing interface context, and adapting assistance to on users' goals and interactions. However, these capabilities have largely been examined separately. Less work has examined how such systems can combine interface and user context to support BLV screen reader users in constructing and adapting mental models of UI.

%% file: 3_positionality.tex
\section{Positionality}
The research team consisted of three researchers with prior experience working with disabled communities; however, none of us identify as BLV. We recognize that our perspectives are shaped by our positionality and that our understanding of BLV experiences is inherently limited. Following standpoint theory~\cite{harding_standpoint_1992}, we engaged two BLV community members as consultants to help center community knowledge, priorities, and daily lived experiences. The first consultant teaches BLV individuals how to use accessibility technologies at a large metropolitan center for the blind. They have no functional vision. The second consultant is an accessibility leadership advocate who works with organizations to improve disability inclusion and accessibility. They have light perception but no functional vision. The first author met with both consultants over four sessions to discuss project scope, emerging insights, and solicit feedback on study design. Consultants also tested \systemname{} in preparation for the user evaluation study. We compensated consultants at a rate of \$40 per hour-long session for their time and expertise.

%% file: 4_interview.tex
\section{Interview Study on BLV Mental Models of UI}
We conducted a semi-structured interview study to investigate how BLV screen reader users understand and access modern UIs. Specifically, we inquired into how BLV screen readers' constructed, adapted, and used mental models of UIs, particularly when encountering inaccessible, inconsistent, or dynamic interfaces.

\subsection{Participants}
We recruited 11 (5 female, 6 male) BLV participants through BLV-centered organizations and mailing lists (Table~\ref{tab:interview-demographics}). Participants were eligible if they: (1) were at least 18 years old, (2) self-identified as blind or low vision, and (3) had any screen reader experience. Participants received a \$15 USD Amazon or Visa gift card of choice as compensation for their time. 

\newcolumntype{Y}{>{\raggedright\arraybackslash}X}
\begin{table*}[t]\centering
\centering
\caption{Interview Participant Demographics.} 
\Description{This is a table of Interview Participant demographics. This table has 12 rows and 6 columns. The first row is a header row with the following columns: ID, Gender, Age, Vision Level, Age experienced vision loss, and Screen readers used.}
\small
\rowcolors{2}{gray!10}{white}
\begin{tabularx}{\linewidth}{l l l X l X}
\toprule
\textbf{ID} & \textbf{Gender} & \textbf{Age} & \textbf{Vision Level} & \textbf{Age experienced vision loss} & \textbf{Screen readers used}\\
\midrule
I1 & M & 31 & Low Vision (usable vision) & 10 & VoiceOver, NVDA, JAWS \\
I2 & F & 55 & Totally Blind (light perception) & 49 & VoiceOver, NVDA, JAWS \\
I3 & F & 31 & Totally Blind (no light perception) & Birth & VoiceOver, NVDA, JAWS \\
I4 & F & 27 & Low Vision (usable vision) & 23 & VoiceOver \\
I5 & M & 28 & Low Vision (usable vision) & 1 & VoiceOver, JAWS\\
I6 & M & 24 & Totally Blind (no light perception) & 1 & VoiceOver, NVDA, JAWS\\
I7 & M & 57 & Totally Blind (no light perception) & 47 & VoiceOver, NVDA, JAWS\\
I8 & F & 75 & Totally Blind (no light perception) & 1 & VoiceOver, JAWS\\
I9 & M & 59 & Totally Blind (light perception) & 45 & VoiceOver, NVDA, JAWS\\
I10 & M & 74 & Totally Blind (no light perception) & Birth & JAWS\\
I11 & F & 23 & Low Vision (usable vision) & 2 & VoiceOver, NVDA, JAWS\\
\bottomrule
\end{tabularx}
\label{tab:interview-demographics}
\end{table*}

\subsection{Procedure}
We conducted remote semi-structured interviews over Zoom for approximately 30 minutes per participant. Recorded interviews covered participants' screen reader usage, UI understanding strategies, mental models of UIs, expectations of dynamic, inaccessible, and inconsistent interfaces, and reliance on external support (\eg{} assistance from allies or AI tools) (Appendix~\ref{appendix:interview}). 

To reduce priming around the term \textit{mental model}, we first asked participants experience-based questions about how they learned, navigated, and reasoned about interfaces (\eg{} \textit{``When you come across an unfamiliar or new website, what do you do to learn it?''}) Only after these questions did we ask questions directed towards mental models (\eg{} \textit{``What mental model do you have for applications (\ie{} websites and apps)?''}). When participants were unfamiliar with the term \textit{mental model}, we reframed the question, asking how they might visualize or structure an interface in their mind. This helped participants describe their own understanding without needing to adopt our terminology~\cite{mack-cultivating-access-2022}. In analysis, we interpreted mental models through their descriptions of navigation strategies, expectations of interfaces and interactions, and interface reasoning.

Following the interview questions, participants shared their screen and walked the interviewer through a website they had recently used; we wanted to see how participants navigated and described an interface they were familiar with. During this semi-structured task, participants described how what they understood about the website, what aspects were clear or confusing, and additional information they wished they had when initially encountering or in further use. 

\subsection{Analysis}
We used a hybrid inductive--deductive thematic analysis approach~\cite{braun_using_2006} to analyze interview transcripts. The first author coded all transcripts, combining open coding with codes from a shared, evolving codebook to capture participants' experiences, needs, and values (\eg{} \textit{``Context-dependent navigation strategies''} and \textit{``Evaluating success of interaction''}). Concurrently, the second author coded half of the transcripts, and both authors collaboratively developed and refined the codebook as new patterns emerged. 
We analyzed walkthrough recordings using interaction analysis~\cite{derry-interactionanalysis-2010}. The first author reviewed recordings alongside participants' navigation, screen reader audio output, and focus to examine how their described strategies and interpretations appeared during interaction. Walkthrough observations were used to contextualize, rather than independently code, participants' interview accounts.
Through affinity diagramming and collaborative discussion, the research team consolidated codes into higher-level themes~\cite{spall-perr-debriefing-1998, braun_using_2006}.

\section{Interview Findings}



\subsection{Screen Reader Mediation Shapes Mental Models of UI}
\label{sec:screen-reader-mediation-directs-expectations}

\subsubsection{Screen reader mediation shapes the available interface}
\label{sec:screen-reader-mediation-shapes-available}

Participants’ screen readers and navigation strategies shaped which aspects of the interface were available for understanding and interaction. Such strategies were shaped by the information exposed by the interface, prior screen reader experience, and individual preferences.
Headings provided high-level page structure, which I2 described as the primary means through which they oriented themselves within an interface. Others, such as I7, used accessibility landmarks as reference points: \texttt{<header>} indicated the top of an interface and \texttt{<footer>} indicated its bottom. 
When such cues were absent, participants often fell back on ``tabbing''---traversing interface elements one-by-one to ``start from scratch and figure [the interface] out.'' (I9)
Participants also relied on navigation strategies developed through screen reader experience and personal preference: I10 relied on 30 years of learned shortcut keys, whereas I7 did not want to memorize shortcuts and primarily used tabbing. 
Rather than reasoning directly from the interface as designed or visually rendered, participants constructed mental models from the interface through their screen reader: the \textit{manifest interface}~\cite{zhang-interaction-proxies-2017}. The manifest interface therefore shaped which information could inform participants' mental models. Differences in accessibility semantics, screen reader behavior, and navigation strategy could make different aspects of the same underlying interface salient to different users, while serial navigation constrained when and in what context information was encountered~\cite{zhang-interaction-proxies-2017, vigo-snapshot-2014, vigo-coping-2013, reyes-cruz-designing-extend-2021}. 

\subsubsection{Prior knowledge grounds interface expectations}
\label{sec:prior-knowledge-grounds}
Participants drew on prior knowledge to construct mental models and form expectations about interface structure, content, functionality, and behavior. The source and specificity of this knowledge, however, depended on their familiarity with the interface: for familiar interfaces, participants reused interface-specific mental models; for unfamiliar interfaces, they drew on generalized knowledge formed across prior interface experiences.
With familiar interfaces, participants relied on accumulated prior knowledge of the interface and its interactions. I1 characterized their mental model of their desktop as a fixed arrangement where ``[they] are able to click the same place [they] clicked yesterday'' to complete tasks. Repeated experience with the same interface supported increasingly specific expectations~\cite{heinz-expect-it-2017}. However, that specificity also made subsequent changes particularly disruptive: I11 described how iOS 26 moved familiar search fields from the top to the bottom of the screen, while I8 could no longer use a links-based navigation strategy after a familiar website changed its layout. Such changes invalidated both individual expectations and the strategies built around them, requiring participants to determine which parts of their prior understanding remained valid and develop new strategies (I5, I10, I11).
With unfamiliar interfaces, participants could not draw on a prior, interface-specific mental model. Instead, they drew on experiences with similar interfaces, with I5 explicitly describing this process: ``[I] compare [the unfamiliar interface] to similar [interfaces]...that I've used before.'' Such experiences informed abstract schemata---generalized knowledge structures formed through prior interface encounters---that participants used to anticipate likely interface structure, actions, functionality, and behavior~\cite{human_computer_interaction_preece_1994, design_of_everyday_norman_2013, saei_mental_2010, mental_models_hci_carroll_1988}. 
Familiarity shaped the knowledge available for expectation formation. While familiar interfaces supported expectations grounded in concrete, interface-specific mental models, unfamiliar interfaces prompted participants to draw on schemata derived from prior interfaces. This suggests that expectation formation operates at different levels of abstraction: interface-specific knowledge supports relatively specific expectations, while schemata provides broader expectations.

\subsubsection{Expectations guide exploration and interaction}
\label{sec:expectations-guide-exploration}
Participants operationalized their mental models through expectations that directed interface exploration and interaction. Due to the incremental nature of screen readers, these expectations allowed participants to selectively explore relevant interface content and interactions rather than rely on exhaustive serial navigation.
On familiar interfaces, interface-specific expectations helped participants navigate directly to known content:  ``On a familiar site, [...] I'll know exactly what to do to get to where I want to be.'' (I10) For unfamiliar interfaces, broader expectations similarly directed exploration: I6, I8, I9, I10, and I11 all used \textit{JAWS\footnotemark[15] Find} or in-built browser \textit{Find} shortcuts to search for expected keywords in the UI (further discussed in Section~\ref{sec:perceivable-design-patterns}). 
Exploration and interaction, in turn, provided evidence through which participants evaluated interface expectations as aligned, misaligned, or unresolved.
When interface information participants encountered aligned with an expectation, that evidence could reinforce their existing understanding (I1, I5).
However, when information misaligned with an expectation, participants explored further, tried alternative navigation or interaction strategies, or revised their understanding. For example, while walking through Target.com, I11 interpreted a familiar element as a carousel and expected that navigating to the next element would ``swipe to a new'' carousel item. When the interaction instead revealed that the element was now a banner, the mismatch prompted I11 to revise their understanding of how this once familiar element behaved.
In other cases, participants could not readily determine whether an apparent expectation mismatch reflected absent content, unsuccessful or incorrect behavior, inaccessible information, or a differently implemented interface. Such uncertainty made expectation evaluation difficult. While completing dental paperwork, I2 recalled encountering an element they understood to be a button, but the element was not clickable. Unable to determine how the element was intended to behave or how to proceed, they abandoned the form altogether. 
Expectations, exploration, and interaction operate reciprocally: prior knowledge formed initial expectations that guided exploration and interaction, while available interface information provided evidence to reinforce or revise one's mental model and subsequent expectations~\cite{soper-the-nature-2016, norman-observations-1987, liu-considering-2010}. 
However, an expectation mismatch did not necessarily indicate that participants' mental model was incorrect. Because participants evaluated the interface through a screen reader, they could be  uncertain whether content was absent or simply not exposed. They then had to decide whether to revise their understanding, continue exploring, or attribute the mismatch to limitations in how the interface was exposed.

\subsection{Cross-Interface Knowledge Transfer Depends on Perceivable Design Patterns}
\label{sec:design-patterns}

\subsubsection{Perceivable design patterns guide initial expectations}
\label{sec:perceivable-design-patterns}
Participants leveraged perceivable design patterns to form initial expectations about unfamiliar interfaces. Design patterns are general archetypes for addressing an interaction goal~\cite{dearden-pattern-lang-hci-2006, van-welie-interaction-patterns-ui-2000, design_of_everyday_norman_2013}. For example, one common web interface design pattern for supporting navigation is a horizontal navigation bar listing top-level categories, with sub-categories available through pull-down menus. A design pattern can be implemented in many ways, yet each implementation shares core ideas. 
Rather than approaching unfamiliar interfaces as entirely new, participants drew on recurring patterns associated with interface categories (\eg{} e-commerce, social media, particular brands) and interfaces within the same brand family~\cite{roth-mentalmodels-web-objects-2010}. For example, I1 recognized similarities across Meta products such as Facebook and WhatsApp, while I8 noted that shopping websites consistently have a ``search box somewhere,'' whereas investment firm websites lack consistency in their design.
For participants who navigated by searching for expected text, recurring terminology was a cue through which patterns could be recognized:  ``some people call it sign-in, some people call it login [...] if everyone could use the same verbiage [...] it would make things really easy.'' (I11)
Recurring elements and controls also supported pattern recognition, with I5 explaining that consistent in how elements and controls behaved made it ``easier to navigate quickly, and build accurate mental models.'' 
Perceivable design patterns provided participants with initial expectations for UI understanding, allowing them to reuse learned meanings and strategies. When new interfaces followed recognizable patterns, participants could readily apply prior knowledge, reducing the effort required to construct understanding from scratch~\cite{design_of_everyday_norman_2013, mental_models_laird-2010, mental_models_hci_carroll_1988, roth-mentalmodels-web-objects-2010}. Thus, cross-interface knowledge transfer depended not only whether interfaces shared underlying design patterns, but on whether the cues that made those patterns recognizable were available through screen reader interaction. 



\subsubsection{Imperceptible design patterns disrupt transfer}
\label{sec:imperceptible-design-patterns-disrupt}
While perceivable design patterns supported more efficient UI understanding, imperceptible patterns---whether due to inaccessibility or inconsistency---made participants' familiar patterns less useful for transferring knowledge. 
While walking through an e-commerce website, I3 expected a search field but could not locate one: ``there should be [...] a search...I couldn't find a search.'' Although the search field was present, it was not perceivable through I3's screen reader, making the pattern effectively unavailable.
Participants also described cases in which familiar controls were represented differently through screen reader interaction than the visually rendered interface. I9 mentioned how buttons and dropdown menus were often implemented for sighted interaction, and would not interact with their screen reader. 
I6 likewise recalled what was supposedly a button being exposed as ``just text,'' limiting interaction. 
In these cases, an element could visually resemble a familiar pattern while its screen reader semantics did not.
Inaccessibility could obscure defining characteristics such as semantics and behavior, while inconsistent implementation could make otherwise perceivable elements difficult to associate with known patterns. Thus, design patterns were only transferred when their defining characteristics were preserved in the screen-reader-mediated manifest interface.

\subsection{Interface Changes Create Expectation Misalignment and Uncertainty}
\label{sec:interface-changes-create-expectations}

\subsubsection{Interaction-driven changes require verifying expected outcomes}
\label{sec:interaction-driven}
Interaction-driven changes gave participants a causally grounded expectation for what should happen after an action; however, sufficient evidence of the resulting interface state was not always immediately exposed through their screen reader, if at all. 
When small interaction-driven updates were not surfaced, participants had to manually verify whether their action succeeded. For example, after I7 added products to their cart on an e-commerce website, their screen reader remained ``silent'', forcing them to navigate to the cart and verify whether the products were correctly added. Although such a change may have been immediately visible on the page through an updated cart count, it was not exposed through screen reader interaction.
When more disruptive interaction-driven changes occurred, participants recognized that something had changed (\eg{} through a focus shift) but could not easily determine their location or the new interface state. For example, upon navigating through Zoom's login process for the interview, I5 explained that the updated layout and newly introduced elements made the login options difficult to find. 
 In both cases, screen reader mediation made it harder to connect an interaction to its outcome and assess whether the resulting state matched participants' expectations. Small, unsurfaced updates prompted explicit verification, whereas larger changes required participants to reorient themselves and reconstruct their understanding of the interface. Even when an interaction provided a known causal anchor, breakdowns in the interface, screen reader, or relationship between them could obscure whether the expected outcome occurred~\cite{borodin_more_2010, bostic2019exploringintersectionswebscience, carvalho-accessibility-usability-problems-2018, bigham-2017-dontknow, vision_marr_2010, giudice_navigating_2018}. 



\subsubsection{Autonomous changes can remain unnoticed or unexplained} 
\label{sec:autonomous-changes}
Autonomous interface changes (\eg{} an alert added to the top of the interface, an intrusive pop-up alert randomly taking over the interface) occurred without a preceding user action to establish what should happen next.
Participants described both changes that interrupted their interaction and those that remained silent. 
I6 described pop-up alerts as disruptive because they ``kind of [change] the whole [interface]'',  prompting them to reorient themselves. By contrast,
when autonomous updates did not interrupt ongoing interaction, participants became aware of them only through later re-exploration or explicit notice from others (I5, I7). 
These forms of autonomous change created different problems. Disruptive updates signaled that a change had occurred but interrupted participants' flow, leaving them to determine what had changed and how to proceed. Silent updates posed a more fundamental detection problem, leaving participants unaware that the interface had changed at all. Because neither provided an action-based causal anchor, participants lacked a specific action for anticipating the change or interpreting the resulting interface state.

\subsection{BLV Users Synthesize Fragmented Context to Understand UI}
\label{sec:blv-users-synthesize-fragmented-context}



\subsubsection{Alternative sources of context supplement screen reader output}
\label{sec:alternative-sources}

Screen reader output did not always provide all of the context participants needed to understand UI.
Participants therefore supplemented available screen reader information with other devices, tools, people, and interaction modalities that exposed different aspects of the interface.
Across various interfaces, participants described their screen readers to skip and stumble over content (I1, I10), improperly describe content (I1), or jumble content together (I3). When screen reader output was insufficient, participants turned to other sources to reconstruct the interface. For example, I11 described switching from JAWS to VoiceOver on their iPad to use touch-based spatial exploration, perceiving relationships that were absent from their desktop screen reader: ``I can understand it [...] by touching around, and then go back to my laptop and pull it up.'' 
Participants also sought out supplementary information from their communities, remote visual interpreting services, OCR tools, and both general-purpose (\eg{} ChatGPT, Gemini) and specific accessibility AI tools (\eg{} JAWS PictureSmart, Be My AI) to obtain additional interface context~\cite{adnin_i_2024, hayes-cooperative-privacy-2019, feng-understanding-inform-2024, penuela-investigating-2024, sharma_before_2025}. 
In particular, participants valued AI tools because human assistance was not always available or reliable (I1, I9). 
These accounts suggest that participants constructed UI understanding across multiple partial representations, selecting sources according to what information was needed and what assistance was available.

\subsubsection{Visuospatial context provides visual content and spatial relationships}
\label{sec:visuospatial-context}

One especially important gap concerned visual content and spatial relationships. Graphical content could be inaccessible when it lacked sufficient textual alternatives; even when individual elements were accessible, their organization, proximity, and relative position could remain difficult to infer through serial screen reader navigation.
Participants therefore valued visuospatial context that was not readily available through their usual screen reader interaction~\cite{potluri_examining_2021, chheda-kothary_it_2025, giudice_navigating_2018, gubbi_mohanbabu_context-aware_2024}. 
I1 wanted to know what pictures on a news website looked like but was presented with insufficient alternative text. Similarly, I9 encountered button elements implemented as images without alternative text, preventing screen reader access. These examples show that inaccessible visual content could obscure both the meaning of interface information and the functionality of interactive elements.  
Participants also lacked spatial understanding, even when element functionality was accessible, as these relationships are often not surfaced by screen readers (I7, I11)~\cite{chheda-kothary_it_2025, chheda-kothary_understanding_2023}. 
Visuospatial context functioned as additional context for understanding otherwise incomplete interface representations. Visual content clarified what inaccessible or ambiguous content was, while spatial knowledge helped participants understand organization, proximity, and relationships among elements. These accounts suggest that screen-reader-accessible content may be insufficient for UI understanding when important visual contents or spatial relationships remain unavailable.

%% file: 5_design.tex
\section{Design Objectives}
\label{sec:design-objectives}
Based on our interview findings, related work, and prototyping with BLV consultants, we derived five design objectives (DOs) for intelligent systems that support BLV UI understanding. These objectives necessarily serve as hypotheses, informed by insights reported up to this point, about what would improve the quality and efficiency of screen-reader-mediated interaction for UI understanding. We subsequently interrogate these hypotheses by instantiating the design objectives in \systemname{} (Section~\ref{sec:coral}) and evaluating it with BLV screen reader users (Section~\ref{sec:eval}).

\begin{itemize}
    \item[\textbf{DO1}] \textbf{Contextualize the current interface using users' interaction history.} Systems should relate the current interface to users' prior interactions and previously encountered states rather than treating each encounter independently. Repeated use produced interface-specific expectations and navigation strategies, which participants reused to reduce exploration  and interaction effort (Section~\ref{sec:prior-knowledge-grounds}). When interfaces misaligned with participants' expectations, they had to revise or reconstruct portions of their previously established understanding (Sections~\ref{sec:expectations-guide-exploration},~\ref{sec:interaction-driven}). Because conventional screen readers primarily expose the current interface state, systems should maintain relevant prior interactions and interface states, and use them to contextualize the current interface against prior encounters~\cite{gajos07:automatically, gajos_personalized_2012, potluri_examining_2021}.


  

    \item[\textbf{DO2}] \textbf{Make familiar design patterns perceivable and recognizable.} Systems should surface the defining characteristics of familiar design patterns when they are not coherently exposed through screen reader interaction. 
    Experience with recurring patterns contributes to schemata generalized across interfaces, allowing users to form initial interface expectations rather than constructing UI understanding from scratch~\cite{design_of_everyday_norman_2013, mental_models_laird-2010, mental_models_hci_carroll_1988, liu-considering-2010, dearden-pattern-lang-hci-2006, van-welie-interaction-patterns-ui-2000}.
    Participants transferred schemata across interfaces using recurring structure, semantics, terminology, and behavior (Section~\ref{sec:perceivable-design-patterns}), but inaccessible or inconsistent implementations could make a familiar pattern appear absent or different (Section~\ref{sec:imperceptible-design-patterns-disrupt})~\cite{janeiro-semantic-pattern-2010, dearden-pattern-lang-hci-2006, van-welie-interaction-patterns-ui-2000, design_of_everyday_norman_2013}.
    Systems should identify patterns in the underlying interface and make both their defining characteristics and deviations recognizable in the manifest interface.

    
    \item[\textbf{DO3}] \textbf{Make interface changes noticeable and interpretable.}
    Systems should surface meaningful interface changes that are not readily apparent through screen reader interaction and provide  evidence for interpreting the resulting state. 
    Prior work shows that screen reader users can face ambiguity when interface information appears absent, changed, or inaccessible, as they cannot always determine the origin of this ambiguity~\cite{borodin_more_2010, bostic2019exploringintersectionswebscience, bigham-2017-dontknow, carvalho-accessibility-usability-problems-2018}.
    Given this, participants manually verified interaction-driven changes against a preceding action and its expected outcomes, re-explored interfaces when autonomous updates lacked such an action-based anchor, and reconstructed their understanding when familiar interfaces changed over time (Sections~\ref{sec:interaction-driven},~\ref{sec:autonomous-changes}). 
    Systems should indicate what changed and where, relate the updated state to recent actions or prior states when supported by available evidence, and avoid implying causal relationships that cannot be established.
    

    

    \item[\textbf{DO4}] \textbf{Convey interface context lacking in screen reader interaction.} Systems should convey interface context that is difficult to obtain through screen reader interaction alone, including high-level organization, visual semantics, and spatial relationships.    
    Participants assembled this context from partial information provided by sighted assistance, AI tools, and additional screen readers or devices (Section~\ref{sec:alternative-sources})~\cite{potluri_examining_2021, adnin_i_2024, borodin_more_2010}. Additionally, participants used visual semantics to interpret inaccessible or ambiguous content, while spatial information helped them understand broader interface structure (Section~\ref{sec:visuospatial-context})~\cite{chheda-kothary_it_2025, giudice_navigating_2018, chheda-kothary_understanding_2023}.
    Systems should synthesize these complementary forms of context into concise support relevant to users' current goals, reducing the effort required to reconcile partial representations themselves.
    
    \item[\textbf{DO5}] \textbf{Integrate support into established screen reader workflows.} Systems should augment rather than replace users' established screen reader practices.    
    Participants selected navigation strategies according to their experience, efficiency, familiarity, and personal preferences, and incorporating complementary tools when their typical workflows were insufficient (Section~\ref{sec:screen-reader-mediation-shapes-available}). Systems should therefore deliver support through familiar screen reader conventions, preserve users' configurations and learned strategies, and avoid requiring them to adopt a separate interaction model~\cite{reyes-cruz-designing-extend-2021, wimer-kanchi-nonvisual-2026}.
\end{itemize}

Our subsequent evaluation of \systemname{} prompted refinements to \textbf{DO3} and \textbf{DO5}, discussed in Sections~\ref{sec:catch-and-replay} and \ref{sec:coordinating-coral}, respectively. Appendix~\ref{appendix:revised-dos} presents the complete set of these design objectives with these revisions incorporated.

%% file: 5_coral.tex
\section{The \systemname{} System}
\label{sec:coral}
We instantiated these design objectives by developing \systemname{}\footnote{The source code is available at \url{https://anonymous.4open.science/r/AI-Narrations-425C/}}, a context-aware browser extension that generates gists to surface meaningful interface states and changes for BLV screen reader users. To generate these gists, \systemname{} combines \textit{interface context}---structural, semantic, and visual evidence about the current UI---with \textit{user context}, a system-side representation of information that may shape the user's understanding. User context includes recent interactions, prior interface states, and representations of interfaces previously encountered while using \systemname{}; it does not directly represent the user's mental model. Drawing on both forms of context, gists describe the current interface state or change, identify where it occurred, and relate it to recent interactions or prior states when supported by available evidence.
Gists are narrated through the user's screen reader, allowing \systemname{} to work with common screen readers (\eg VoiceOver\footnote{\url{https://support.apple.com/guide/voiceover/welcome/mac}}, NVDA\footnote{\url{https://www.nvaccess.org/download/}}, JAWS\footnote{\url{https://vispero.com/jaws-screen-reader-software/}}). Through these mechanisms, \systemname{} is designed to help users in maintain an understanding of the interface and interpret interface changes. \systemname{} is designed for Chromium-based browsers and implemented in TypeScript and React. 

\subsection{Usage Scenario}
Consider Greg, a screen reader user who uses \systemname{} alongside his screen reader. Greg visits an new electronics e-commerce website to purchase a ``Nova X15'' laptop. The website reflects accessibility barriers common across the web: some elements and controls lack appropriate programmatic semantics, while dynamic changes are not consistently conveyed to screen readers~\cite{bi_accessibility_2022, martin-large-scale-2024, webaim_million, fok_large_scale_2022}. Based on prior experiences with e-commerce interfaces, Greg expects the website to follow familiar design patterns and contain components such as a grid of product listings, filters that narrow the product listings, and a cart that updates as items are added. However, he does not yet know which patterns the website follows, how its components are organized, or whether they will be consistently accessible via his screen reader.
When the page loads, \systemname{} plays a subtle ticking sound, providing \textit{progress feedback} that it has detected and is interpreting the new interface state. Once processing is complete, \systemname{} narrates Gist~U1: 

\gisttable
 {U1}
 {Initial Site Gist}
 {UI Action}
 {Initial Load}
 {The website features a header, a product filter sidebar, and a main content area showing a grid of Electronics, similar to familiar online store layouts.}

Gist~U1 summarizes the page's high-level organization and relates it to familiar e-commerce patterns from \systemname{}'s \textit{Experience Store}. This provides Greg an initial basis for directing his exploration without first traversing the page serially.
Greg remembers that \systemname{} identified a product-filter region. Using his usual screen reader commands, he navigates by heading and finds ``Product Filters.'' Greg sets the brand filter to ``Nova,'' expecting the product listings to update. However, focus remains on the filter, and his screen reader provides no indication of whether the expected update occurred. After providing progress feedback, \systemname{} relates the filter selection to the resulting interface change and narrates Gist~U2.

\gisttable
 {U2}
 {Products Update Gist}
 {User Action}
 {Brand Filter Set to ``Nova''}
 {The product grid has updated to display only the Nova X13 and Nova X15 laptops.}

 Gist~U2 provides evidence that the interaction succeeded and identifies the resulting product, including the ``Nova X15.'' When Greg briefly forgets the product name, he uses the \textit{Replay} command to hear the gist again. He then uses his screen reader's \textit{Find} command to move directly to the Nova X15 and open its product page. Here, \systemname{} informs rather than replaces Greg's established navigation strategies.
Upon encountering the product page, \systemname{} provides progress feedback while interpreting the new state. Because elements such as the navigation bar persisted across the page transition, \systemname{} does not repeat them in Gist~U3. Instead, the gist describes how the new state differs from the preceding page.

\gisttable
 {U3}
 {Product Page Gist}
 {UI Action}
 {Initial Load}
 {The product detail page for the X15 has loaded, removing the filters sidebar and displaying specifications, an image, and an ``Add to Cart'' button.}
 
Greg finds and activates ``Add to Cart.'' His screen reader provides no indication that the cart was successfully updated; however, \systemname{} relates the interaction to the resulting change in the cart state and narrates Gist~U4. 

 \gisttable
 {U4}
{Cart Update Gist}
 {User Action}
{Activated ``Add to Cart''}
 {The cart link in the top navigation bar has been updated to show one item after adding the product to your cart.}

The gist confirms that the Nova X15 was added and that the cart now contains one item, allowing Greg to evaluate the outcome without navigating to the cart solely for verification. Greg opens the cart and proceeds to checkout.
Shortly after the checkout page loads, while Greg is navigating near the bottom of the page, the interface autonomously adds a required shipping-information region near the top. Because this region appears earlier in the page's navigation order and was not caused by Greg's most recent action, he has no reason to return to the beginning of the interface. Without additional feedback, he might discover the change only after being unable to place the order and re-exploring the page. \systemname{} provides progress feedback that something has changed and then narrates Gist~U5. 

 \gisttable
 {U5}
 {Shipping Information Gist}
 {UI Action}
  {Autonomous Interface Update}
 {The checkout page has loaded with shipping options, while the site header and navigation bar are no longer visible.}

 Greg locates the new region, completes the required fields, and places his order.
Throughout these interactions, \systemname{} supplements rather than replaces Greg's existing screen reader workflow. Its gists provide contextual evidence with which he can form expectations about an unfamiliar interface, evaluate whether interactions produced their expected outcomes, and revise his understanding as the interface changes.

\subsection{Pipeline Overview}

\begin{figure*}
 \centering
 \includegraphics[width=0.99\linewidth,alt={Diagram for the Coral pipeline made up of three stages: UI Observation, Context Modeling, and Gist Generation. During UI Observation, the Document Object Model is observed by Interaction Listeners and a Mutation Observer which creates a Page Payload. The Page Payload is passed to Context Modeling, where Interface Context is generated. Interface Context is made up of Structural, Semantic, and Visual Context. Additionally, User Context is retrieved, made up of the Experience Store and Interface State and Interaction Histories. After Context Modeling, Importance Evaluation is run using State Comparison, Text Delta Handling, and LLM Evaluation. If the change is important, the gist is generated and narrated during the Gist Generation stage where Semantic Context, State, and Interactions are saved. Users can also use Coral Replay to narrate the gist again.}]{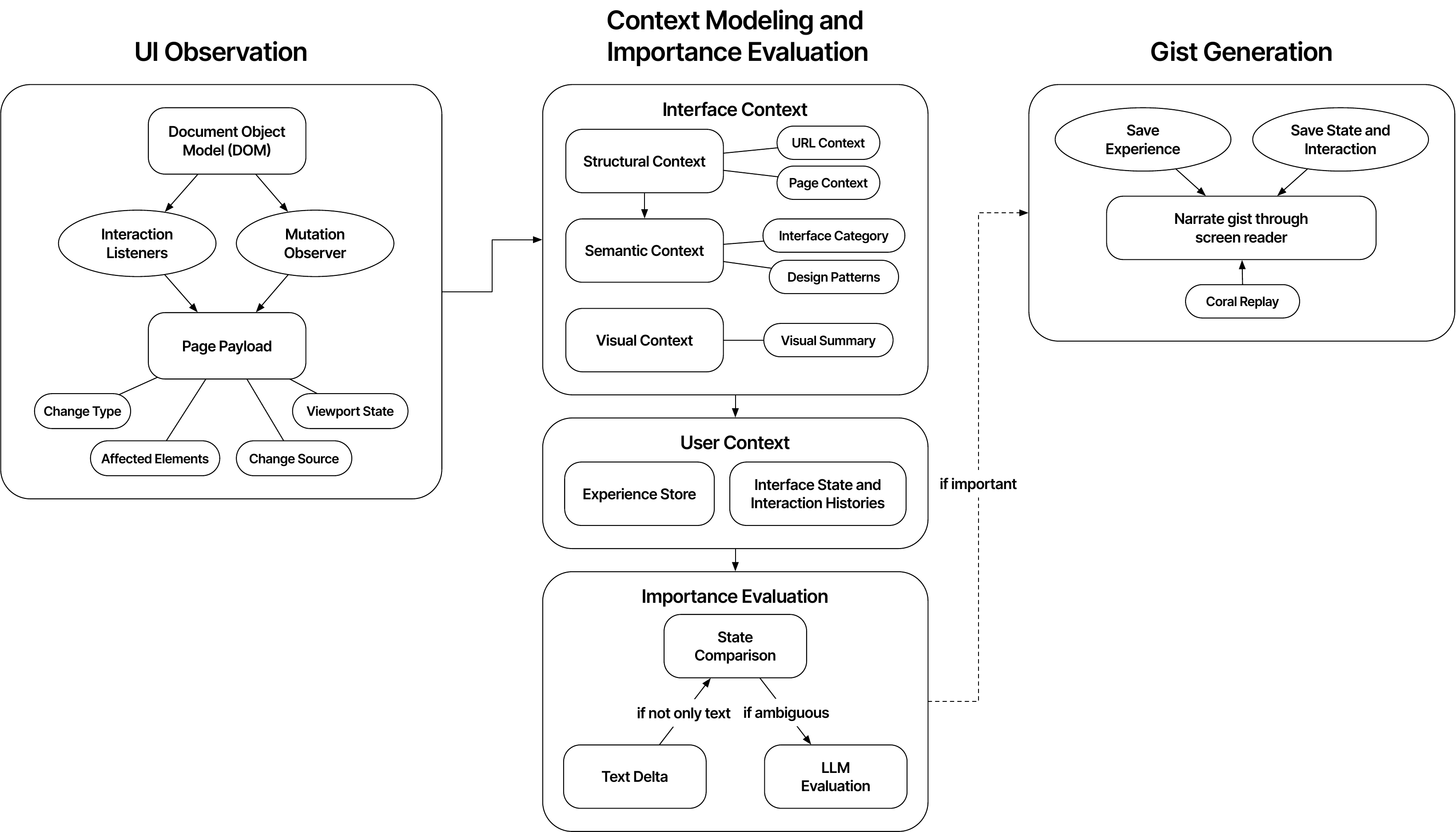}
 \caption{The \systemname{} multi-stage pipeline, comprised of (1) UI Observation, (2) Context Modeling and Importance Evaluation, and (3) Gist Generation.}
 \Description{Diagram for the Coral pipeline made up of three stages: UI Observation, Context Modeling, and Gist Generation. During UI Observation, the Document Object Model is observed by Interaction Listeners and a Mutation Observer which creates a Page Payload. The Page Payload is passed to Context Modeling, where Interface Context is generated. Interface Context is made up of Structural, Semantic, and Visual Context. Additionally, User Context is retrieved, made up of the Experience Store and Interface State and Interaction Histories. After Context Modeling, Importance Evaluation is run using State Comparison, Text Delta Handling, and LLM Evaluation. If the change is important, the gist is generated and narrated during the Gist Generation stage where Semantic Context, State, and Interactions are saved. Users can also use Coral Replay to narrate the gist again.}
 \label{fig:coral-pipeline}
\end{figure*}

\systemname{} operates as a multi-stage pipeline (Figure~\ref{fig:coral-pipeline}): (1) \textbf{UI Observation}, (2) \textbf{Context Modeling and Importance Evaluation}, and (3) \textbf{Gist Generation}. UI Observation converts detected interface mutations and recent user interactions into a payload of what changed, where it occurred, and its possible relationship to those interactions. Context Modeling uses this payload to construct interface context and retrieve relevant user context, while Importance Evaluation determines whether the change is sufficiently meaningful to surface. When a change is considered important, Gist Generation synthesizes the both forms of context into a concise narration. 
Throughout this process, \systemname{} records user interactions and updates the interface-state baseline used to interpret subsequent changes. After generating a gist, \systemname{} stores the gist and its semantic representation in the \textit{Experience Store}, where they can be retrieved for retrieval during future interactions. 
Table~\ref{tab:coral-traceability} maps each objective to the components that operationalize it.

\newcolumntype{Y}{>{\raggedright\arraybackslash}X}

\begin{table}[t]
\centering
\caption{\systemname{} features mapped to design objectives.}
\Description{This is a table of Coral features mapped to design objectives. This table has 14 rows and 2 columns. The first row is a header row with the following columns: Feature and Design Objective(s). The third and seventh rows have subheaders with a full row.}
\small
\renewcommand{\arraystretch}{1.25}
\rowcolors{2}{gray!10}{white}
\begin{tabularx}{\linewidth}{Y l}
\toprule
\textbf{Feature} & \textbf{Design Objective(s)}\\
\midrule
UI Observation (Section~\ref{sec:ui-observe}) & DO3\\
\midrule
\multicolumn{2}{l}{\textit{\textbf{Interface Context}}} \\
\midrule
Structural Context (Section~\ref{sec:structural-context}) & DO4\\
Semantic Context (Section~\ref{sec:semantic-context}) & DO2, DO4\\
Visual Context (Section~\ref{sec:visual-context}) & DO4\\
\midrule
\multicolumn{2}{l}{\textit{\textbf{User Context}}} \\
\midrule
Interface State and User Interaction History (Section~\ref{sec:interaction-context}) & DO1, DO3\\
Experience Store (Section~\ref{sec:interaction-context}) & DO2\\
\midrule
Importance Evaluation (Section~\ref{sec:importance}) & DO3\\

Gist Generation (Section~\ref{sec:gist}) & DO1, DO2, DO3, DO4\\

Screen Reader Narration (Section~\ref{sec:sr-injection}) & DO5\\

Progress Feedback (Section~\ref{sec:progress}) & DO5\\

\systemname{} Replay (Section~\ref{sec:replay}) & DO5\\

\bottomrule
\end{tabularx}
\label{tab:coral-traceability}
\end{table}

\subsection{UI Observation}
\label{sec:ui-observe}
\systemname{} observes initial page loads and continuously monitors the DOM for subsequent structural and textual changes caused by user interactions or autonomously. A mutation observer tracks added, removed, and updated nodes, while interaction listeners record recent clicks and significant key presses (\eg{} \texttt{Enter}, \texttt{Space}). Interaction events provide evidence for associating subsequent changes with recent user actions; when no interaction is present, \systemname{} treats the change as potentially autonomous.
For each set of detected changes, \systemname{} constructs a \textit{page payload}. The payload contains the affected nodes (\eg{} text nodes and interface elements), change type (\eg{} node removal, text mutation, initial page load), recent user interactions, inferred change source (\eg{} user interaction, autonomously), and page metadata. This payload serves as shared input to \textbf{Context Modeling and Importance Evaluation}.


\subsection{Context Modeling}
\label{sec:context-model}
Context Modeling augments the page payload with two forms of context. \textit{Interface context} comprises complementary structural, semantic, and visual representations of the UI: structural context captures its organization, semantic context captures its purpose and recognizable design patterns, and visual context captures visuospatial information. \textit{User context} situates the current interface in relation to recent interactions, prior interface states, and previously encountered interfaces. Together, these forms inform whether an interface state or change should be surfaced and determine what information a gist should contain.

\subsubsection{Structural Context}
\label{sec:structural-context}
Structural context represents the interface's high-level organization using DOM-derived information. It captures structure commonly exposed through screen reader navigation with a compact representation of how the interface is organized. This allows downstream stages to reason about page structure without processing the full DOM.

\paragraph{Page Context.}
Page context is extracted during UI Observation (Section~\ref{sec:ui-observe}) and reflects the current interface state. It includes page metadata (\eg{} page title, description, URL), selected text content, headings, and landmarks. \systemname{} also adapts \citet{surfers2005}'s summarization algorithm to produce a concise summary of the page's content.

\paragraph{URL Context.}
URL context is a cached, model-generated description of the interface's structure. Using the Gemini\footnote{We use Google's \href{https://ai.google.dev/gemini-api/docs/models/gemini-3-flash-preview}{Gemini 3 Flash} model for all LLM and VLM calls; we selected this model in particular for its speed, reduced latency, and input/output size (1M/65K tokens). Queries are made with the Thinking level set to Minimal. The only exception to this is during gist generation, when the change is not only text-based. We use institutional access to this model, with guarantees and safeguards to avoid model training upon user data (further discussed in Section~\ref{sec:limitations}).} API and Prompt~\ref{appendix:prompt-url}, \systemname{} generates a description focusing on likely page regions and interface organization rather than visual presentation. The resulting description is associated with the URL and reused across subsequent interface changes, reducing the need to regenerate page-level structural context.

\subsubsection{Semantic Context}
\label{sec:semantic-context}
Semantic context represents the interface's broad purpose and reusable design patterns. Using structural context, \systemname{} generates a category (\eg{} \textit{e-commerce}, \textit{dashboard}) with keywords describing recurring layout, navigation, and interaction design patterns (\eg{} \textit{header-navigation}, \textit{product-grid}, \textit{sidebar-filtering}). We use Prompt~\ref{appendix:prompt-semantic} to generate a category and keywords that summarize the interface's primary function, patterns, and behavior. Because these  patterns often remain stable across incremental interface updates, \systemname{} reuses semantic context unless state comparison indicates a substantial change (Section~\ref{sec:importance}). The resulting category and keywords are used in gist generation and as the searchable representation stored in the Experience Store (Section~\ref{sec:interaction-context}).



\subsubsection{Visual Context}
\label{sec:visual-context}
Visual context captures visual interface content that may not be fully represented in the DOM. \systemname{} captures a screenshot of the current viewport and provides it to the Gemini\footnotemark[14] image understanding API. While visual context is limited to the current viewport, DOM observation and structural context operate across the page, observing content outside the user's current viewport. Using Prompt~\ref{appendix:visual}, the model generates a summary of visible content,  grouping, layout, and interactive regions. This representation complements structural context when spatial relationships, visibility changes, or visually presented interface regions may not be exposed through screen reader interaction.

\subsubsection{User Context}
\label{sec:interaction-context}
User context is a system-side representation of information that may shape the user's understanding. It comprises recent interactions, prior interface states, and representations of interfaces previously encountered while using \systemname{}. This context allows \systemname{} to relate the current interface state to recent user actions and identify recurring design patterns across encountered interfaces. It does not directly represent the user's internal mental model, familiarity, or expectations; rather, it provides user traces from which \systemname{} can infer relationships relevant to gist generation.


\paragraph{Interface State and User Interaction History}
\systemname{} maintains a local record of recent interface states and user interactions. Interactions are pruned every minute, while interface states persist in the database. When a change immediately follows a recorded action, this history provides evidence that the change may be an outcome of that interaction. It also allows \systemname{} to compare the current state with prior states and avoid describing updates as isolated events.


\paragraph{Experience Store}
The Experience Store supports retrieval across previously encountered interfaces for design pattern recognition. After generating a gist, \systemname{} stores the gist together with its semantic category and pattern keywords. When a new interface is encountered, \systemname{} performs semantic retrieval\footnote{We use the OpenAI \href{https://developers.openai.com/api/docs/guides/retrieval}{Retrieval API} for our vector stores and related querying functionality. We also use institutional access to this model, because the Experience Store contains contains representations of participants' interactions and previously encountered interfaces, to safeguard against model training on their data (Section~\ref{sec:limitations}).}, using the current interface's semantic category and pattern keywords, retrieving up to three related interfaces. Retrieved experiences are supplied to gist generation as optional context for describing similar or different design patterns.


\subsection{Importance Evaluation}
\label{sec:importance}
For each detected change, \systemname{} determines whether the change should be surfaced and what contextual information should be regenerated to interpret it. 
\systemname{} first compares the current and prior states, then handles text-only changes and deterministic noise filtering. Changes with unresolved importance proceed to LLM Evaluation.
Changes that do not meaningfully alter the interface state or affect users' understanding are ignored, while meaningful changes proceed to Gist Generation.


\subsubsection{State Comparison}
\systemname{} constructs a partial representation of the current interface state from the page payload and interface context---comprising of semantic and structural context---and compares it to the previous state representation. This comparison occurs before visual and user context are generated. Based on the comparison, \systemname{} categorizes the current state as the \textit{same}, \textit{similar}, or \textit{different}, determining whether the detected change proceeds through the pipeline and how much interface context should be regenerated. 
States are categorized as the \textit{same} when their metadata, semantic context, and selected content match; \systemname{} treats the detected mutations as redundant and does not surface the change. States are categorized as \textit{similar} when they differ in keywords or selected content; \systemname{} reuses stable contextual representations and regenerates only those affected by the change. States are categorized as \textit{different} if the page metadata differs; \systemname{} regenerates the full interface context. After comparison, the current partial representation becomes the baseline for evaluating the next change, regardless of whether the current change proceeds to gist generation.

\subsubsection{Text Delta}
When a change consists only of text content, \systemname{} prioritizes the explicit text delta and reuses interface context. For example, if a label changes from ``Cart (1)'' to ``Cart (2)'', the updated text can be interpreted relative to existing interface context. This allows \systemname{} to preserve existing context and avoid model calls for otherwise stable interfaces.

\subsubsection{Noise Filtering}
\systemname{} applies deterministic rules to remove changes unlikely to affect UI understanding. These include duplicate mutations, minor CSS-property updates, and background changes that do not alter functionality or user-relevant content. Changes to \texttt{display} and \texttt{visibility} properties are retained as they may affect available interactive regions. Additionally, mutations involving semantically important regions (\eg{} \texttt{<main>}, \texttt{<section>}, \texttt{<dialog>}) are prioritized for further evaluation.

\subsubsection{LLM Evaluation}
When deterministic mechanisms cannot establish if a change is meaningful, \systemname{} queries the Gemini\footnotemark[14] API using Prompt~\ref{appendix:importance}. The model receives the page payload, interface context, and preceding interface state and returns a boolean importance decision. A change is considered important when it affects a user's interaction, communicates the outcome of a recent interaction, or introduces content likely to affect the user's understanding. Changes classified as unimportant do not proceed in the pipeline.

\subsection{Gist Generation and Interactions}
\label{sec:gist}
Once a state or change is deemed important, \systemname{} generates a one-sentence gist using Prompt~\ref{appendix:gist}. The prompt receives (1) the current page payload, (2) interface context, and (3) relevant user context.
The resulting gist describes what changed and where and, when supported by available evidence, relates the resulting user interface state to a recent user interaction or previously encountered interface Gists use plain language to describe the current state, with retrieved interactions and interface patterns helping prioritize content relevant to the user's ongoing interaction.

Each generation is associated with the interface state from which it was produced; if subsequent changes occur before the generation completes, \systemname{} queues the outstanding gist.
After gist generation, \systemname{} stores the gist and its associated semantic context in the Experience Store.


\subsubsection{Screen Reader Narration}
\label{sec:sr-injection}
 To integrate with users' existing screen reader workflows, \systemname{} proactively narrates gists through the user's screen reader rather than through a separate text-to-speech system. This preserves existing speech settings, including rate, pitch, and verbosity. When a gist is generated, \systemname{} inserts it into a visually hidden element with \texttt{aria-live="assertive"}\footnotemark[10], causing compatible browser–screen reader configurations to announce the update.  We tested this mechanism with VoiceOver\footnotemark[11], NVDA\footnotemark[12], and JAWS\footnotemark[13]. 

\subsubsection{Progress Feedback}
\label{sec:progress}
Gist generation may require several seconds depending the interface context, user context, and model latency. When \systemname{} detects a potential change and begins processing it, the system plays a subtle ticking sound. The sound stops when narration begins or when the change is determined as unimportant.
This feedback distinguishes change detection from completed interpretation: users are informed that \systemname{} is processing an updated without interrupting the screen reader's current speech. This feedback is produced through the browser's audio system rather than the screen reader speech queue, allowing users to continue navigating during generation.

\subsubsection{\systemname{} Replay}
\label{sec:replay}
Since \systemname{} operates asynchronously and proactively, users may continue navigating while a gist is being generated or narrated. Subsequent screen reader output may interrupt or replace the gist, reflecting standard screen reader behavior. To support gist recovery, \systemname{}'s Replay feature allows users to revisit a missed or interrupted gist through a command that fits alongside existing screen reader workflows. To use Replay, users can activate a default keyboard shortcut (\texttt{Ctrl/Command+Shift+Y}) that replays the most recently generated gist. 

%% file: 6_evaluation.tex
\section{User Evaluation}
\label{sec:eval}


To evaluate both our design objectives (Section~\ref{sec:design-objectives}) and whether \systemname{} supported them in practice, we conducted a user evaluation with BLV screen reader users completing structured tasks on dynamic web interfaces under baseline and \systemname{} conditions. We used \systemname{} as a research probe rather than treating task performance as the primary outcome. 
Following both conditions, reflective interviews captured participants' experiences, preferences, and frictions when using \systemname{} compared to baseline screen reader interaction.


\subsection{Participants}
We recruited 8 BLV participants through BLV-centered organizations and mailing lists (Table~\ref{tab:eval-demographics}). Participants were eligible if they: (1) were at least 18 years old, (2) self-identified as blind or low vision, (3) frequently used common screen readers (\eg{} VoiceOver, NVDA, JAWS), and (4) were familiar with their desktop file manager for \systemname{} installation. Participants received a \$25 USD Amazon or Visa gift card of choice as compensation for their time. 

\newcolumntype{Y}{>{\raggedright\arraybackslash}X}
\begin{table*}[t]\centering
\centering
\caption{Evaluation Participant Demographics.} 
\small
\Description{This is a table of Evaluation Participant Demographics. This table has 9 rows and 8 columns. The first row is a header row with the following columns: ID, Gender, Age, Vision Level, Age experienced vision loss, Screen readers used, Baseline, and Coral.}
\rowcolors{2}{gray!10}{white}
\begin{tabularx}{\linewidth}{l l l X l X | l l}

\toprule
\textbf{ID} & \textbf{Gender} & \textbf{Age} & \textbf{Vision Level} & \textbf{Age experienced vision loss} & \textbf{Screen reader used} & \textbf{Baseline} & \textbf{\systemname{}}\\
\midrule
P1 & M & 56 & Totally Blind (no light perception) & 30 & JAWS & Site B & Site A\\
P2 & M & 42 & Totally Blind (light perception) & Birth & JAWS & Site A & Site B\\
P3 & M & 39 & Totally Blind (no light perception) & 4 & JAWS & Site B & Site A\\
P4 & M & 81 & Totally Blind (no light perception) & 35 & JAWS & Site A & Site B\\
P5 & M & 28 & Low Vision (usable vision) & 1 & JAWS & Site B & Site A\\
P6 & M & 64 & Totally Blind (no light perception) & 22 & JAWS & Site A & Site B\\
P7 & M & 46 & Totally Blind (light perception) & Birth & VoiceOver & Site B & Site A\\
P8 & M & 36 & Totally Blind (light perception) & 14 & VoiceOver & Site A & Site B\\
\bottomrule
\end{tabularx}
\label{tab:eval-demographics}
\end{table*}

\subsection{Procedure}
\label{sec:eval-procedure}
We conducted a remote, within-subject task-based evaluation lasting approximately 60 minutes per participant. Participants joined over Zoom using their own computers and shared their screens. They completed tasks under two conditions: (1) a baseline condition using only their screen reader and (2) an experimental condition using \systemname{} alongside their screen reader. Throughout the session, participants were asked to think aloud by verbalizing their thoughts, expectations, reactions, and challenges. This allowed us to examine how they understood the interface, evaluated interface changes, and made sense of \systemname{} and its gists. \systemname{} was installed and removed during the session, either by participants with guidance or by the interviewer through permitted remote control~\cite{mack-cultivating-access-2022}. To preserve participant-specific user context, each participant used a personal build of \systemname{} with a unique Experience Store (Section~\ref{sec:interaction-context}).

To mitigate content familiarity effects, we developed two mock e-commerce websites that were functionally isomorphic but whose interfaces appeared substantially different. We counterbalanced website assignment across conditions. We chose e-commerce websites because these interfaces commonly include complex layouts, varied interaction patterns, and retrieval-heavy tasks~\cite{yu_cluttered_2025}. Table~\ref{tab:site-differences} summarizes the differences between the two websites. Each participant completed the baseline condition on one website and the \systemname{} condition on the other.
Condition order was fixed, with baseline preceding \systemname{}. This design is occasionally used in behavioral research~\cite{ma2025chatgpt,trinh2014pitchperfect,wright2018corticospinal} and it is considered appropriate when the baseline condition reflects the ``business-as-usual'' design that participants are already familiar with, and where there are concerns that the intervention condition might introduce skills or strategies that participants might carry over to subsequent interactions. 

After each condition, participants described their understanding of the website and answered 5-point Likert-scale questions assessing UI understanding (Appendix~\ref{appendix:evaluation}). As in the interview study, when participants were unfamiliar with the term \textit{mental model}, we asked how they understood or organized the website in their mind~\cite{mack-cultivating-access-2022}. After the \systemname{} condition, participants also completed the System Usability Scale (SUS)~\cite{brooke_sus_1996, bangor2009determining}. Following both conditions, participants took part in a semi-structured exit interview focused primarily on their experiences with \systemname{}, using the baseline condition as a point of comparison. The interview covered participants' UI understanding, expectations, cognitive effort, value and utility, and speculative use of \systemname{}. 

\newcolumntype{C}{>{\centering\arraybackslash}X}
\begin{table}[t]
\centering
\caption{Differences between the mock e-commerce websites.} 
\Description{This is a table of the differences between Mock Websites. This table has 6 rows and 2 columns. The first row is a header row with the following columns: Site A and Site B.}
\small
\rowcolors{2}{gray!10}{white}
\begin{tabularx}{\linewidth}{C C}
\toprule
\textbf{Site A} & \textbf{Site B} \\
\midrule
Grid of Product Listings & Table of Product Rows \\
Filter Sidebar & Refine Dialog Menu \\
No Navigation Bar on Checkout & Navigation Bar on Checkout \\
Cart Page & Cart Sidebar \\
Table of Cart Item Rows & List of Cart Items \\
\bottomrule
\end{tabularx}
\label{tab:site-differences}
\end{table}

\subsubsection{Tasks}
 We did not treat task completion alone as evidence of UI understanding. Instead, the tasks were designed to elicit moments in which participants formed expectations, evaluated whether interface behavior aligned with those expectations, and, when necessary, revised or reconstructed their understanding of the interface. 
 Across tasks, participants navigated between pages and interface regions, located relevant information and controls, and encountered interface changes. The tasks progressively introduced additional interaction demands: establishing initial shopping cart state (T1), differentiating and adding a second product (T2), recalling and modifying prior state (T3), and selecting options and confirming task completion (T4) (Table~\ref{tab:site-tasks}).
 Together, these tasks provided structured opportunities to observe how participants maintained UI understanding during ongoing interaction. 

\newcolumntype{Y}{>{\raggedright\arraybackslash}X}
\begin{table}[t]\centering
\centering
\caption{User Evaluation Tasks.} 
\Description{This is a table of the User Evaluation Tasks. This table has 5 rows and 4 columns. The first row is a header row with the following columns: Task \#, Site A Tasks, Site B Tasks, and Task Goals.}
\small
\rowcolors{2}{gray!10}{white}
\begin{tabularx}{\linewidth}{l X X X}
\toprule
\textbf{Task \#} & \textbf{Site A Tasks} & \textbf{Site B Tasks} & \textbf{Task Goals}\\
\midrule
T1 &
Add a product with 16GB of RAM to the cart twice &
Add a product with 256GB of Storage to the cart &
Locate product based on attributes and establish cart state\\

T2 & 
Add a Nova-branded product (different from the first product) to the cart &
Add a Atlas-branded product (different from the first product) to the cart &
Differentiate products and extend the existing cart state\\

T3 & 
Reduce the quantity of the first product to 1 in your cart &
Increase the quantity of the second product to 3 in your cart &
Recall and modify a previously established cart state\\

T4 & 
Checkout with Standard Shipping &
Checkout with Express Shipping  &
Locate and select an option, then confirm task completion\\
\bottomrule
\end{tabularx}
\label{tab:site-tasks}
\end{table}


\newcolumntype{Y}{>{\raggedright\arraybackslash}X}
\begin{table*}[t]\centering
\centering
\caption{\systemname{} System Usability Scale. Responses were collected on a 5-point scale (\ie{} 1 = Strongly disagree; 5 = Strongly agree). Questions suffixed with * are reverse coded (\ie{} 1 = Strongly agree; 5=Strongly disagree). Overall scores of at least 73 are interpreted to mean ``good'' usability and scores of 85 or more are interpreted as ``excellent''~\cite{bangor2009determining}.} 
\small
\Description{This is a table of the Coral System Usability Scale results. This table has 12 rows and 9 columns. The first row is a header row with the following columns: Question, P1, P2, P3, P4, P5, P6, P7, and P8. The last row is a footer row with the Overall SUS score.}
\rowcolors{2}{gray!10}{white}
\begin{tabularx}{\linewidth}{X l l l l l l l l}
\toprule
\textbf{Question} & \textbf{P1} & \textbf{P2} & \textbf{P3} & \textbf{P4} & \textbf{P5} & \textbf{P6} & \textbf{P7} & \textbf{P8} \\
\midrule
I think I would like to use Coral frequently & 3 & 5 & 5 & 5 & 3 & 5 & 5 & 2 \\
I found Coral unnecessarily complex* & 5 & 5 & 4 & 4 & 5 & 5 & 5 & 5 \\
I thought that Coral was easy to use & 5 & 5 & 4 & 4 & 4 & 5 & 5 & 5 \\
I think that I would need the support of a technical person to be able to use Coral* & 5 & 5 & 5 & 5 & 5 & 5 & 5 & 5 \\
I found the various functions in Coral were well integrated & 5 & 5 & 4 & 4 & 4 & 5 & 5 & 5 \\
I thought there was too much inconsistency with Coral* & 3 & 5 & 5 & 4 & 4 & 4 & 5 & 5 \\
I would imagine that most people would learn to use Coral very quickly & 5 & 5 & 4 & 4 & 4 & 5 & 5 & 5 \\
I found Coral very cumbersome to use* & 5 & 5 & 4 & 5 & 5 & 5 & 5 & 5 \\
I felt very confident using Coral & 5 & 5 & 4 & 5 & 4 & 5 & 5 & 5 \\
I needed to learn a lot of things before I could get going with Coral* & 5 & 5 & 5 & 4 & 5 & 4 & 5 & 5 \\
\specialrule{1pt}{0pt}{2pt}
\textbf{Overall SUS score (possible range 0--100)} & \textbf{90} & \textbf{100} & \textbf{85} & \textbf{85} & \textbf{82.5} & \textbf{95} & \textbf{100} & \textbf{92.5} \\
\bottomrule
\end{tabularx}
\label{tab:coral-sus}
\end{table*}

\subsection{Analysis}
We used a hybrid inductive--deductive thematic analysis approach~\cite{braun_using_2006} to analyze session transcripts. We also used interaction analysis~\cite{derry-interactionanalysis-2010} to examine participants' screen reader and \systemname{} use across conditions. The first and second authors reviewed each recording alongside participants' screen reader output and focus, examining how participants responded to \systemname{}'s progress feedback and narration, used screen reader functionality with and without \systemname{}, and adapted their navigation strategies when using \systemname{}. 
The first author coded all sessions, combining open coding with codes from a shared, evolving codebook to capture participants' experiences, interactions, needs, and reactions across the baseline and \systemname{} conditions (\eg{}\textit{``Using \systemname{} Replay to verify change''} and \textit{``Unsure about benefit of visual description''}). The second author coded half of the sessions,  and both authors collaboratively refined the codebook as new patterns emerged. 

Through affinity diagramming and discussion, the research team consolidated codes into higher-level themes, considering both participants' experiences with \systemname{} and their relevance to the design objectives~\cite{spall-perr-debriefing-1998, braun_using_2006}. 
To complement the qualitative analyses, we present summary statistics for the SUS and the post-condition Likert-scale responses. Given the small sample size, we treat these analyses as exploratory and interpret the results cautiously alongside participants' observed interactions and qualitative accounts.


\begin{figure*}
 \centering
 \includegraphics[width=1\linewidth,alt={Grouped bar chart showing the average Likert-scale measures for Baseline vs. Coral Conditions. Questions are along the X axis, with each question having two bars, a blue bar for Baseline and a red bar for Coral. The Y axis goes from 1 to 5. Participants were able to understand changes without manual re-explore under the Coral condition by a mean difference of 1.37. All other questions show positive trends, but lack a significant mean difference for statistical significance. For both conditions, all 8 participants understood the functionality of the website, having average measures of 5.}]{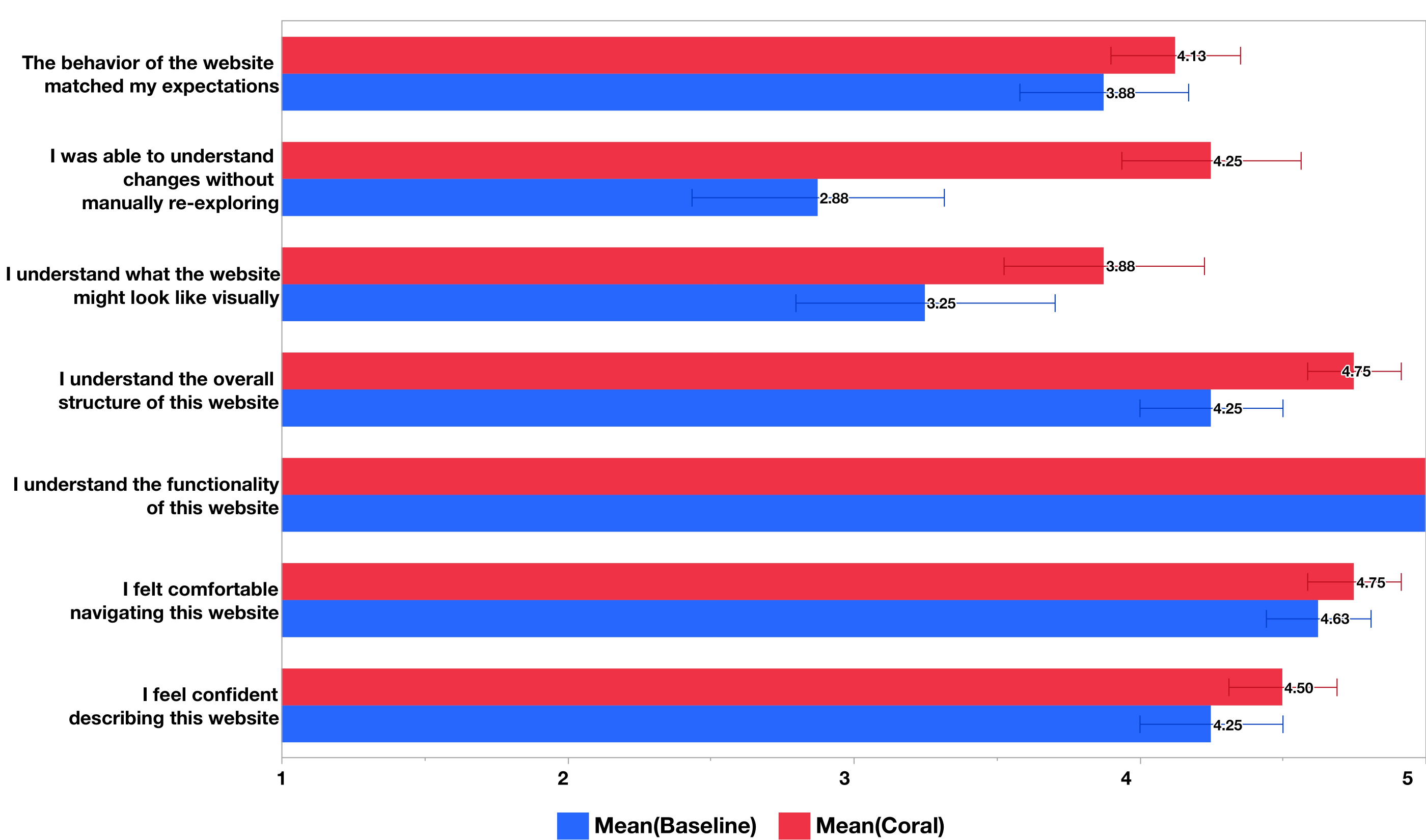}
 \caption{Average Likert-scale Measures for Baseline vs. \systemname{} Conditions}
 \Description{Grouped bar chart showing the average Likert-scale measures for Baseline vs. Coral Conditions. Questions are along the X axis, with each question having two bars, a blue bar for Baseline and a red bar for Coral. The Y axis goes from 1 to 5. Participants were able to understand changes without manual re-explore under the Coral condition by a mean difference of 1.37. All other questions show positive trends, but lack a significant mean difference for statistical significance. For both conditions, all 8 participants understood the functionality of the website, having average measures of 5.}
 \label{fig:coral-likert}
\end{figure*}

\section{User Evaluation Findings}
Across all participants, the mean SUS score for \systemname{} was $91.25 \pm 6.81$ (range $82.5$--$100$), which is generally interpreted as ``excellent'' usability~\cite{bangor2009determining} (Table~\ref{tab:coral-sus}). We interpret this score cautiously; our small participant pool limits generalizability, and SUS may overestimate perceived usability due to ceiling effects or social desirability, especially in a moderated study setting. Additionally, SUS includes alternating positively and negatively worded statements, which can introduce interpretation burden or misunderstandings; however, participants often re-confirmed the scale and statement wording during completion. We therefore do not treat the SUS score as conclusive evidence of usability, but as one descriptive measure interpreted alongside qualitative feedback, including participants' critiques around \systemname{}'s timing, proactivity, and gist value. 
Below, we expand upon these findings based on observed interactions and participant reflections.


\subsection{Constructing Mental Models of Unfamiliar Interfaces}
\subsubsection{Grounding initial expectations with interface-specific context}
\label{sec:grounding-initial}
During initial exploration, participants used \systemname{}'s gists as interface-specific evidence for forming expectations about unfamiliar interfaces.
When P1 encountered Site A with \systemname{}, they found the initial gist to be helpful while they were ``getting used to'' the unfamiliar interface (Gist~1).
Similarly, when \systemname{} described Site A's home page as a ``products page'', P3 drew on their knowledge of e-commerce websites to ``quickly know some important things'' without exploring the entire page: ``I just developed a mental model based on what [\systemname{}] provided.''
Rather than providing a complete mental model, \systemname{} supplied information that participants interpreted through  schemata to form initial expectations and mental model of the interface.

\gisttable
 {1}
 {Initial Site A Gist (P1)}
 {UI Action}
 {Initial Load}
 {Northstar Electronics products page features a top navigation bar, a filters sidebar, and a main content area displaying a grid of items like the Nova X13 and Orbit Air.}

\subsubsection{Synthesizing high-level interface structure}
\label{sec:synthesizing-high-level}
\systemname{} synthesized interface elements and regions into high-level structural summaries that participants used to direct subsequent exploration.
 During the baseline condition, participants constructed this understanding incrementally, often describing the interface by serially enumerating elements they encountered: ``It appears to have a main navigation [...] There is the banner section, where there is apparently a logo, and there are a couple of links.'' (P2).
With \systemname{}, gists communicated multiple interface regions and their organization without requiring participants to explore the entire page. On the checkout page, P3 found \systemname{} helpful for understanding the page as a series of top-to-bottom sections (Gist~2).
 Survey responses reflected the same pattern: the mean rating for understanding the interface's overall structure increased from $4.25$ in the baseline condition to $4.75$ with \systemname{}, a mean difference of $0.50$.

 \gisttable
 {2}
 {Checkout Page (P3)}
 {UI Action}
 {Load of Checkout Page}
 {The checkout page has loaded, removing the top navigation and header to focus on a new content area containing a cart summary, shipping options, and a submit order button.}

\subsubsection{Conveying visual contents and spatial relationships}
\label{sec:conveying-visual-spatial}
\systemname{} provided visuospatial relationships that are often unavailable or difficult to infer through the screen-reader-mediated manifest interface.
Participants appreciated how \systemname{}'s gists incorporated aspects of visual presentation without ``regurgitating exactly what's actually on the screen.'' (P1) 
On Site B, \systemname{} informed P2 that the underlying content was hidden while the filter dialog was open, helping them understand both the interface's current state and which content was interactive (Gist~3). 
Survey responses showed a similar pattern: the mean rating for visual understanding increasing from $3.25$ in the baseline condition to $3.88$ with \systemname{}, a mean difference of $0.63$.
Participants also used \systemname{}'s spatial descriptions to infer implicit relationships. On Site A with \systemname{}, both P3 and P5 noted how the filters were positioned on the left, with P5 reasoning that ``because it said the filters [were] on the left, it automatically means that the products were on the right.'' After using \systemname{}, several participants described the interface in terms of spatial relationships among its regions and elements (P2, P3, P5, P6).


\gisttable
 {3}
 {Filter Dialog (P2)}
 {User Action}
 {Opened ``Refine'' dialog menu}
 {A ``Refine'' menu has opened, providing filter options for brand, RAM, and storage while the catalog content and navigation are hidden.}

\subsection{Evaluating Expectations Across Interface Changes}

\subsubsection{Progress feedback establishes change awareness}
\label{sec:progress-awareness}
All participants valued \systemname{}'s progress feedback as an  immediate indication that the system detected an interface change before narrating its interpretation.
Participants described how the feedback---or, for P1 and P7, ``chirping''---told them that ``something's changed'' (P2, P6). 
P2 explained that the feedback saved them time they would otherwise spend manually re-exploring the page to determine whether a change occurred. 
Even, P8, who we observed to use \systemname{} the least, valued the feedback as a notification: ``it could have just made any kind of sound, [to notify that]...something [...] happened.''
This pattern was consistent with survey responses: on average, \systemname{} led to a large increase in participants' reported understanding of changes without manually re-exploring (mean difference = $1.375$; despite the small sample size, this result was also statistically significant: Wilcoxon signed-rank, $p=0.0156$). Although this measure reflects the \systemname{} condition as a whole, participants' accounts identified progress feedback as supporting immediate change awareness.
These findings distinguish awareness that the interface changed from interpretation of what changed and why. 
Progress feedback indicated that participants' current understanding of the interface state might not align, while the subsequent gist supported interpretation. 
Participants further operationalized this distinction through \textit{Catch-and-Replay}, which we further discuss in Section~\ref{sec:catch-and-replay}.


\subsubsection{Connecting interaction-driven changes to expected outcomes}
\label{sec:connecting-changes}
Interaction-driven changes provided participants with expectations about the resulting interface state. 
In the baseline condition, participants often had to seek evidence that these expected outcomes had occurred. On Site B during Task T1, P1 said ``I have one item in my cart, so therefore it must be in there,'' navigating through cart elements to verify the added item.
With \systemname{}, participants spent less time and reported less effort evaluating interaction outcomes, with some not manually verifying at all (P1, P2, P3, P4, P7). After P2 added an item to the cart, \systemname{} notified them that the cart was updated to have one item. P2 explained that this saved the time otherwise required  ``to go to [cart] and find out'' whether the interaction succeeded (Gist~4). By connecting the preceding interaction, \systemname{} could describe the resulting interface state in relation to participants' actions. This transformed an ambiguous interface change into evidence of whether the expected outcome had occurred, reducing the need for manual verification.

\gisttable
 {4}
 {Adding to cart (P2)}
 {User Action}
 {Added Nova X13 laptop to cart}
 {The cart button in the top navigation bar has updated to show one item added.}

\subsubsection{Relating interface states to prior states}
\label{sec:relating-interface-states}
\systemname{} also contextualized the current interface by comparing it with previously encountered states, providing continuity across updates and page transitions.
During page transitions, P1 noted how \systemname{} related the new page to the previous one: ``it actually told me what was different. [It] told me like what [...] page [I was] transitioning from, and where I was transitioning to. And then what was actually changing, what was different? So that was pretty [...] intuitive.'' . 
After P5 adjusted product filters for Task T1, \systemname{} indicated that only the filter area and products had changed, rather than suggesting that the entire page had been updated (Gist~5). 
 

\gisttable
 {5}
 {Filter Changes (P5)}
 {User Action}
 {Setting the Brand filter to ``Nova'' and RAM filter to ``8GB''}
 {The filters area has been updated with selectable options for Brand, RAM, and Storage, and the product list now shows the Nova X13.}




\subsubsection{Incomplete evidence leaves expectations unresolved}
\label{sec:incomplete-evidence}
However, \systemname{} did not always provide the information participants needed to clearly evaluate their expectations, leading them to revert to manual verification.
After filtering products, \systemname{} did not report the updated product count, so P2 navigated to Site B's table to verify the number of displayed rows. During Task T3, almost every participant wanted \systemname{} to explicitly state that the product quantity had increased or decreased. Instead, \systemname{} reported the updated subtotal, requiring participants to infer whether their interaction had succeeded; even after making this inference, participants wanted the causal outcome to be stated directly (Gist~6). Thus, expectation evaluation depended not only on whether \systemname{} surfaced a change, but also on whether the system supplied the specific evidence participants needed to interpret the expected outcome.

\gisttable
 {6}
 {Updating Product Quantity (P6)}
 {User Action}
 {Increasing the quantity of the Atlas Go laptop from 2 to 3}
 {The Cart Summary side panel has been updated to show an Atlas Go laptop and a new subtotal of \$2836.}

\subsection{Integrating \systemname{} into Established Screen Reader Workflows}

\subsubsection{Coordinating \systemname{} with screen reader interaction}
\label{sec:coordinating-coral}
Although \systemname{} narrated gists through participants' existing screen reader configurations, participants still had to learn how to coordinate its proactive output with their ongoing interaction. Participants appreciated not needing to adopt a separate speech system; P1, for example, said that they ``didn't have to get too used to anything else speaking''. However, using the same speech source was less distinguishable, and P1 found it occasionally difficult to differentiate their screen reader from \systemname{}. 
\systemname{}'s asynchronous timing introduced an additional coordination challenge.
During early use, most participants paused their navigation while \systemname{}'s progress feedback played, waiting for the gist to be narrated. 
This waiting could interrupt their workflow; P6 described being ``a little impatient, waiting for the [feedback] to hear what it was saying.'' As sessions progressed, however, participants began incorporating \systemname{} into their existing navigation strategies. P7  explained, ``I could hear the [feedback] and realize, okay, it's working. I can still move around a bit, but I'm ready for it to start [narrating]''. 
Using \systemname{} alongside a screen reader involved more than preserving existing configurations: participants gradually learned its behavior and how to coordinate its support with ongoing navigation.
These findings refine \textbf{DO5: Integrate support into established screen reader workflows} by distinguishing technical integration from integration into users' workflows. Integrating support requires not only preserving existing configurations, but making system output distinguishable and enabling users to coordinate its timing with their interaction.

\subsubsection{Controlling interface interpretation through Catch-and-Replay}
\label{sec:catch-and-replay}

Many participants used the the \systemname{} Replay shortcut to repeat gists; for example, P6 used it to double-check their understanding of the interface. More unexpectedly, several participants developed a new interaction paradigm that we call \textit{``Catch-and-Replay''}. Upon hearing the progress feedback, they silenced an incoming gist by immediately navigating and replayed it when they were ready to hear it. 
P4 used this strategy the most, explaining: ``I had the ability to just ignore [the gist] and go around to the next thing I wanted to check and then have it speak when I wanted it to.'' 
Building on the change awareness described in Section~\ref{sec:progress-awareness}, \textit{Catch-and-Replay} allowed participants to separate the immediate indication that the interface had changed while deferring \systemname{}'s interpretation until it fit their interaction needs. 
This behavior refines \textbf{DO3: Make interface changes noticeable and interpretable}; noticeability and interpretability do not need to occur simultaneously. Systems can first provide a lightweight notification that a change occurred, then allow users to access evidence about what changed, where it occurred, and how it related to them when they are ready to interpret it.

\subsection{Understanding \systemname{}'s Perceived Value and Utility}

\subsubsection{Using \systemname{} can introduce interpretive and attentional effort}
When first using \systemname{}, several participants described additional interpretative effort as they learned how to understand and use its information. For some, this effort decreased as they became more familiar with \systemname{}. P3, for example, said that their effort initially increased because they ``[had] to try to figure out what...[the] information that [\systemname{}] provides actually means...'' 
These demands varied across participants and their interaction strategies. P4 reported no additional effort because \textit{Catch-and-Replay} allowed them to control when \systemname{}'s narration played. 
In contrast, P8, experienced ongoing attentional effort from having to filter the progress feedback and gists from their own thoughts. While \systemname{} reduced some of the effort required to understand UI, it could introduce new demands associated with interpreting and attending to its support. 

\subsubsection{Relevance and grounding shape \systemname{} reliance}
\label{sec:relevance-grounding}
Participants' ability to utilize \systemname{}'s gists depended on whether the gist was sufficiently specific for their goals, interpretable through existing workflows, and grounded in the manifest interface. 
P8, for example, wanted more explicit spatial detail, such whether elements were ``on the top left.''
However, participants did not uniformly value greater visuospatial specificity. P3, who had been totally blind with no light perception for a most of their life, was concerned that \systemname{} may favor visual content over the structure exposed through their screen reader, noting ``the disconnect between the way screen readers experience websites and the way people see.''
Generated language could also shape participants' navigation even when it was not grounded in the interface. During Task T4, \systemname{} referred to ``shipping method options,'' leading P1 to search for ``method,'' although that term did not appear on the page. P1 nevertheless valued the gist for providing language to support search-based navigation. This case highlights a tension in synthesized support: generated descriptions can scaffold participants' existing strategies, but ungrounded terminology could also create misaligned expectations and additional work. Reliance on \systemname{} depended not simply on receiving additional information, but on whether that information addressed participants' goals while remaining interpretable and grounded in the interface. 

\subsubsection{Value depends on task context and user preferences}
\label{sec:value-depends}
Participants envisioned \systemname{} as most valuable when uncertainty and exploration effort were high. 
A majority wanted to use it on complex and unfamiliar interfaces (P1, P2, P5, P7, P8). P2 explained that \systemname{} could be ``super helpful [when] there is a lot to find out,'' particularly because it was ''watching out for changes.'' 
In contrast, participants perceived less value when \systemname{} repeated information they already understood or confirmed expected interface states  (P3, P6, P8). During Task T3, P8 emptied their cart while believing they were increasing a product's quantity. Although \systemname{} reported the resulting change, P8 ignored the gist because ``a lot of things [\systemname{}] told me were things I already expected and kind of knew was gonna happen.'' P8 therefore did not use the gist and maintained misaligned understanding. This case illustrates how perceived redundancy could reduce attention to \systemname{}, even when a gist contained information that could align the user's current understanding.
 Participants proposed several forms of customization and personalization. P6 suggesting varying levels of verbosity, P3 emphasized that \systemname{} should account for differences in how individuals use interfaces, and P2 wanted it to contextually prioritize information directly related to the preceding action: ``if I increase the amount of elements in the cart, I want to know if that has increased. I may not need to know more.'' 
Providing additional information was not inherently or always valuable; \systemname{} was most useful when its timing, verbosity, and content addressed participants' current uncertainty while supporting their goals and strategies.

%% file: 7_new_discussion.tex
\section{Discussion}
Together, with our interview study and user evaluation of \systemname{}, we discuss how intelligent systems can support BLV screen reader users' understanding of UI by: (1) supplementing element-level access with holistic interface context, (2) situating interface information in user context, (3) supporting design pattern transfer across semantic variation, and (4) separating awareness of interface changes from their interpretation. 

\subsection{Supplementing Element-Level Access with Holistic Interface Context}
Prior work highlighted how BLV screen reader users often construct UI understanding through effortful serial navigation of individual elements, requiring them to integrate information as they explore the UI~\cite{potluri_examining_2021,bigham-2017-dontknow, borodin_more_2010}. Even when individual elements are accessible, screen reader interaction may not communicate how they relate to larger regions, are rendered visually or spatially, or reflect the current interface state~\cite{giudice_navigating_2018,potluri_examining_2021,chheda-kothary_it_2025}. When such information is unavailable, the relationships users can perceive and reason about through the manifest interface are limited~\cite{zhang-interaction-proxies-2017}. While element-level accessibility is necessary, it does not necessarily make the interface available as a coherent whole for efficient or effective UI understanding.  
Our interview findings illustrate how participants supplemented element-level access with their own prior knowledge and strategies. Preferred navigation strategies helped participants develop a high-level understanding of the interface structure, while expectations guided what they sought and how they interpreted available evidence (Sections~\ref{sec:screen-reader-mediation-shapes-available},~\ref{sec:prior-knowledge-grounds},~\ref{sec:expectations-guide-exploration}). When usable cues were available, these practices allowed participants to selectively explore the interface. 
However, while these practices reduced exhaustive exploration, they relied on relevant evidence being exposed through the screen-reader-mediated manifest interface. When needed evidence was unavailable,
participants sought complementary context from alternative sources (Section~\ref{sec:alternative-sources}). Constructing holistic UI understanding therefore entailed not only accessing individual elements, but synthesizing information across elements and partial representations of the interface.
Building on prior work showing that broader interface context can improve the quality of AI-generated image descriptions and generated guidance~\cite{gubbi_mohanbabu_context-aware_2024, chen-struggle-2026}, \systemname{} synthesized interface context comprising high-level structural organization,  visuospatial information, and changes in interface state to notify users of relevant interface states and changes.
During initial encounters with unfamiliar interfaces, participants interpreted these gists through their prior knowledge to form initial expectations and high-level understandings of the interface without first exploring it in full (Sections~\ref{sec:grounding-initial},~\ref{sec:synthesizing-high-level}).
Participants also used \systemname{}'s visuospatial context to interpret what content visually represented and how interface regions related to each other (Section~\ref{sec:conveying-visual-spatial}). 
Future intelligent systems should supplement element-level access with synthesized holistic interface context while grounding their support in evidence from the underlying UI. 
Making such context more accessible alongside element-level screen reader access can broaden the evidence users use to construct  UI understanding and evaluate interface expectations while preserving existing strategies.







\subsection{Situating Interface Information in User Context}
UI understanding depends not not only on the context available in the interface, but also on how that context relates to users' goals, prior knowledge, and ongoing interaction. Prior work described how interaction histories, user traces, and encountered UI states can externalize relationships across navigation, adapt to user context, and enrich UI descriptions~\cite{wexelblat_footprints_1999, gajos07:automatically, potluri_examining_2021, gajos_personalized_2012}. 
Our interview findings illustrate how relevant user context operates across different temporal scales. On unfamiliar interfaces, participants drew on schemata generalized across prior experiences; on familiar interfaces, they relied on interface-specific mental models developed through accumulated use (Section~\ref{sec:prior-knowledge-grounds}). These forms of prior knowledge supported expectations at different levels of specificity, which participants used to guide exploration and interaction.
\systemname{} operationalized aspects of user context through the \textit{Experience Store}---a record of  design patterns from previously encountered interfaces---and shorter-term histories of recent interface states and interactions. When \systemname{} characterized an unfamiliar interface as a ``products page'', participants could relate that information to prior experiences with similar interfaces (Section~\ref{sec:grounding-initial}).
Recent interaction history also allowed \systemname{} to relate interface changes to preceding actions, providing participants with evidence that reduced or eliminated manual verification while evaluating interaction outcomes (Section~\ref{sec:connecting-changes}).
Comparisons with prior interface states similarly helped participants  distinguish changed content from persisted content following both within-page updates and page transitions (Section~\ref{sec:relating-interface-states}). \citet{potluri_examining_2021} suggested comparing new interfaces to familiar interfaces to produce richer semantic descriptions; our findings suggest that comparison can occur within an interface, using prior states and actions to contextualize its current state.
However, more context did not always provide more useful support. Participants differed in the information they needed and how much interpretation they wanted; \systemname{} sometimes omitted necessary evidence or repeated information participants already understood (Sections~\ref{sec:incomplete-evidence},~\ref{sec:value-depends}). 
Future intelligent systems should treat user context as complementary evidence for situating relevant support. Recent actions and interface states can indicate what may be relevant to users' ongoing interaction, while accumulated experience can identify patterns users have previously encountered.
By using such inferences to report synthesized interface information based on users' likely goals and knowledge, intelligent systems can provide contextual support grounded in what users understand about an interface.










\subsection{Supporting Design Pattern Transfer Across Semantic Variation}
Design patterns support knowledge transfer when their defining characteristics remain recognizable across interfaces. For BLV screen reader users, recurring terminology can serve as a recognizable cue to a familiar pattern; however, interfaces may use different terms to express the same functional concept~\cite{janeiro-semantic-pattern-2010, design_of_everyday_norman_2013,van-welie-interaction-patterns-ui-2000,dearden-pattern-lang-hci-2006}.

Our interview findings showed that participants used expected terminology as a navigation anchor, searching for likely terms to locate content and evaluate whether anticipated elements were present (Sections~\ref{sec:expectations-guide-exploration},~\ref{sec:perceivable-design-patterns}). For example, controls labeled ``login'' and ``sign-in'' both instantiate a familiar authentication pattern, but this equivalence may not be apparent through navigation strategies that depend on matching exact interface text. Such variation may also create barriers for users who are unfamiliar with alternative terms, making a familiar pattern difficult to recognize.
Inconsistent terminology could otherwise make familiar patterns difficult to recognize through screen reader interaction. (Section~\ref{sec:imperceptible-design-patterns-disrupt}). Design pattern transfer depends partly on users being able to connect an interface's terminology to a familiar functional concept. 
Our evaluation of \systemname{} exposed a similar tension. For instance, \systemname{} described ``shipping method options'', prompting a participant to search for the term ``method'', although the term was not present (Section~\ref{sec:relevance-grounding}). ``Shipping methods'' and ``shipping options'' are semantically related, but the generated abstraction was not grounded in the interface's searchable terminology. Synthesized semantics could help users recognize familiar patterns while creating false anchors when the language does not correspond to the manifest interface.
Future intelligent systems could mediate between familiar functional concepts and interface-specific terminology. Semantic search, for example, could treat elements that represent ``login'' and ``sign-in'' as equivalent while still identifying the exact label used in the interface. Generated descriptions could similarly identify a control labeled \textit{Login} as an instance of a familiar sign-in pattern. \citet{janeiro-semantic-pattern-2010} described typed semantic relationships---including equivalence, similarity, enhancement, and use---for connecting UI design patterns and improving their retrieval. Extending these ideas to screen reader interaction, intelligent systems could communicate both the familiar functional concept and its interface-specific expression. This could support design pattern transfer while preserving the exact terminology users need for navigation and verification.


\subsection{Separating Dynamic Change Awareness from Interpretation}
Accessibility mechanisms such as ARIA\footnotemark[4] were designed to expose dynamic interface changes to assistive technologies. However, missing, incomplete, or incorrect implementations remain common, and the presence of ARIA does not ensure that changes are communicated coherently through a screen reader~\cite{bi_accessibility_2022, zhong_screenaudit_2025, martin-large-scale-2024, webaim_million, fok_large_scale_2022}. When changes are poorly communicated, screen reader users may remain unaware that the interface changed or lack the evidence as for why, resulting in misaligned understanding~\cite{bostic2019exploringintersectionswebscience, bigham-2017-dontknow}. While change awareness establishes that the interface changed, change interpretation explains what changed, why, and how the resulting state affects users' ongoing interaction.
Our interview study shows how the need for such support varies with the source of a change. Interaction-driven changes followed a user action, providing a causal anchor from which participants could form expectations about an outcome., However, when the outcome was not communicated, participants still needed to verify it manually (Section~\ref{sec:interaction-driven}). Autonomous changes lacked a causal anchor and could remain unnoticed until participants encountered them through later exploration or received explicit notice(Section~\ref{sec:autonomous-changes}). Changes to familiar interfaces could invalidate previously reliable expectations and navigation strategies, requiring participants to determine which parts of their understanding remained valid (Section~\ref{sec:prior-knowledge-grounds}).
Our evaluation of \systemname{} showed how awareness and interpretation do not need to be simultaneous. Progress feedback provided an immediate indication that \systemname{} had detected and was processing a change. Participants could then continue navigating to silence the resulting gist, and use \textit{Catch-and-Replay} to hear the interpretation if needed (Sections~\ref{sec:progress},~\ref{sec:catch-and-replay}). This strategy separated immediate change awareness from \systemname{}'s proactive explanation, allowing participants to defer the gist, prioritize screen reader feedback from their ongoing interaction, and play the explanation when ready.
Future intelligent systems should distinguish indicating a detected change from explaining it. A lightweight cue can provide immediate awareness, while a user-controlled explanation can communicate needed information. The balance of such control may depend on the source of the change. Awareness may be sufficient when an interaction produces an expected outcome. However, unexpected outcomes, autonomous changes, and changes that invalidate prior knowledge of a familiar interface may require more explicit and immediate interpretation. Systems should allow users to ignore, defer, or revisit explanations while prioritizing interpretation when a change lacks a clear causal anchor or may significantly affect ongoing interaction.







%% file: 8_limitations.tex
\section{Limitations, User Privacy, and Future Work}
\label{sec:limitations}


We evaluated \systemname{} on controlled mock websites rather than real-world UIs with inconsistent accessibility, complex structures, and frequent or overlapping changes. Its ability to distinguishing meaningful changes from insignificant background activity in such settings remains an open challenge. 
Moreover, \systemname{}'s LLM- and VLM-based interpretations are subject to hallucinations, omissions, and inaccurate grounding (as seen in Section~\ref{sec:relevance-grounding}), while model-processing delays sometimes competed with participants' ongoing interaction. 
Such errors may provide misleading or untimely evidence about the interface and consequently affect users' UI understanding.
\systemname{} also introduces privacy risks through its use of third-party model APIs, event listeners, and interaction histories, which may expose sensitive interface content or prior activity, particularly for BLV users~\cite{sharma_before_2025, stangl_dump_2023}. Although institutional safeguards state that data is not retained for training, the extent of these protections remain uncertain. We limited the evaluation to mock websites and installed and removed \systemname{} during each session. Future deployments should investigate use of on-device processing and user control over collected data.
Our evaluation included eight participants with screen reader familiarity and experience navigating desktop file managers. This small sample size limited the range of experiences and interactions were could observe; quantitative measures should therefore be interpreted descriptively and alongside participants' observed interactions and verbal accounts.
The evaluation used two functionally isomorphic e-commerce websites to provide comparable interaction opportunities across conditions. 
Although website assignment was counterbalanced, shared content and task structure may have introduced cross-condition learning effects, allowing participants to transfer practice from one condition to the other. 
Structured tasks may also have  
directed participants toward task completion rather than open-ended exploration and encouraged 
narrower and more deliberate interaction than everyday interface use.
Finally, the hour-long evaluation limited our ability to observe any form of long-term learning or changes in reliance on \systemname{}.
Each participant used \systemname{} on only one mock website, limiting our evaluation of \textbf{DO2: Make familiar design patterns perceivable and recognizable}, which relies on \systemname{} being used 
across unfamiliar interfaces over time. The study also included only interaction-driven interface changes, excluding autonomous behavior---which are difficult to experimentally implement. Future longitudinal and in-the-wild studies should evaluate \systemname{} across varied dynamic and complex interfaces, examine how users recognize and transfer design patterns across interfaces, and determine how reliably it identifies and communicates meaningful autonomous changes. 




%% file: 9_conclusion.tex
\section{Conclusion} 
We conducted an interview study with 11 BLV screen reader users examining how they understand and interact with modern UIs. From these findings, related work, and prototyping with BLV consultants, we derived five design objectives and instantiated them in \systemname{}, a context-aware browser extension that generates contextual, screen-reader-narrated gists. In our user evaluation with 8 BLV screen reader users, participants used \systemname{} to form initial expectations, interpret interface changes, and reduce manual verification of interaction outcomes. Their experiences also exposed limitations in gist relevance and grounding and in coordinating \systemname{} with ongoing screen reader interaction. Together, our findings illustrate how intelligent systems can combine interface and user context to support BLV UI understanding.

\section{Use of AI Disclosure}
The authors used generative AI tools to support debugging-oriented system development tasks. Additionally, generative AI tools were used to assist with proofreading and revising in parts of the manuscript text. All outputs were reviewed and verified by the authors.

%% file: appendix.tex
\section{Revised Design Objectives}
\label{appendix:revised-dos}

\begin{itemize}
    \item[\textbf{DO1}] \textbf{Contextualize the current interface using users' interaction history.} Systems should relate the current interface to users' prior interactions and previously encountered states rather than treating each encounter independently. Repeated use produced interface-specific expectations and navigation strategies, which participants reused to reduce exploration effort (Section~\ref{sec:prior-knowledge-grounds}). When interfaces misaligned with participants' expectations, they had to revise or reconstruct portions of their previously established understanding (Sections~\ref{sec:expectations-guide-exploration},~\ref{sec:interaction-driven}). Because conventional screen readers primarily expose the current interface state, systems should maintain relevant prior interactions and interface states, and use them to contextualize the current interface against prior encounters~\cite{gajos07:automatically, gajos_personalized_2012, potluri_examining_2021}.

     \item[\textbf{DO2}] \textbf{Make familiar design patterns perceivable and recognizable.} Systems should surface the defining characteristics of familiar design patterns when they are not coherently exposed through screen reader interaction. 
    Experience with recurring patterns contributes to schemata generalized across interfaces, allowing users to form initial interface expectations rather than constructing UI understanding from scratch~\cite{design_of_everyday_norman_2013, mental_models_laird-2010, mental_models_hci_carroll_1988, liu-considering-2010, dearden-pattern-lang-hci-2006, van-welie-interaction-patterns-ui-2000}.
    Participants transferred schemata across interfaces using recurring structure, semantics, terminology, and behavior (Section~\ref{sec:perceivable-design-patterns}), but inaccessible or inconsistent implementations could make a familiar pattern appear absent or different (Section~\ref{sec:imperceptible-design-patterns-disrupt})~\cite{janeiro-semantic-pattern-2010, dearden-pattern-lang-hci-2006, van-welie-interaction-patterns-ui-2000, design_of_everyday_norman_2013}.
    Systems should identify patterns in the underlying interface and make both their defining characteristics and deviations recognizable in the manifest interface.
    
    \item[\textbf{DO3}] \textbf{Make interface changes noticeable and interpretable.}
    Systems should surface meaningful interface changes that are not readily apparent through screen reader interaction and provide evidence for interpreting the resulting state. Notice and interpretation should be decoupled, allowing users to receive timely indications of change while controlling when and how much interpretive detail is provided.
    Prior work shows that screen reader users can face ambiguity when interface information appears absent, changed, or inaccessible, as they cannot always determine  the origin of this ambiguity~\cite{borodin_more_2010, bostic2019exploringintersectionswebscience, bigham-2017-dontknow, carvalho-accessibility-usability-problems-2018}.
    Given this, participants manually verified interaction-driven changes against a preceding action and its expected outcomes, re-explored interfaces when autonomous updates lacked such an action-based anchor, and reconstructed their understanding when familiar interfaces changed over time (Sections~\ref{sec:interaction-driven},~\ref{sec:autonomous-changes}). During the user evaluation, participants valued \systemname{}'s progress feedback but sometimes wanted to  defer gist narration when that feedback was sufficient to evaluate an interaction's outcome (Section~\ref{sec:progress-awareness},\ref{sec:catch-and-replay}).
    Systems should indicate what changed and where, relate the updated state to recent actions or prior states when supported by available evidence, and avoid implying causal relationships that cannot be established.
    
    \item[\textbf{DO4}] \textbf{Convey interface context lacking in screen reader interaction.} Systems should convey interface context that is difficult to obtain through screen reader interaction alone, including high-level organization, visual semantics, and spatial relationships.    
    Participants assembled this context from partial information provided by sighted assistance, AI tools, and additional screen readers or devices (Section~\ref{sec:alternative-sources})~\cite{potluri_examining_2021, adnin_i_2024, borodin_more_2010}. Additionally, participants used visual semantics to interpret inaccessible or ambiguous content, while spatial information helped them understand broader interface structure (Section~\ref{sec:visuospatial-context})~\cite{chheda-kothary_it_2025, giudice_navigating_2018, chheda-kothary_understanding_2023}.
    Systems should synthesize these complementary forms of context into concise support relevant to users' current goals, reducing the effort required to reconcile partial representations themselves.
    
    \item[\textbf{DO5}] \textbf{Integrate support into established screen reader workflows.} Systems should augment rather than replace users' established screen reader practices, while ensuring that support is distinguishable from screen reader output and coordinated with users' ongoing interaction.
    Participants selected navigation strategies according to their experience, efficiency, familiarity, and personal preferences, and incorporating complementary tools when their typical workflows were insufficient (Section~\ref{sec:screen-reader-mediation-shapes-available}). During the user evaluation, participants sometimes found \systemname{}'s narration difficult to distinguish from other screen reader output. Waiting for gists to finish could also interrupt their navigation and reduce the efficiency of their established workflows.
    Systems should therefore deliver support through familiar screen reader conventions while signaling its source, preserve users' configurations and learned strategies, and avoid requiring them to adopt a separate interaction model~\cite{reyes-cruz-designing-extend-2021, wimer-kanchi-nonvisual-2026}.
\end{itemize}

\section{Interview Study Questions}
\label{appendix:interview}
\subsection{Semi-structured Interview Questions}
\begin{enumerate}
    \item \textbf{Screen reader usage, preferences, and limitations }
    \begin{itemize}
           \item You mention you used [screen reader]; how long have you been using it?
   \begin{itemize}
     \item What do you like about it?
     \item What do you wish was better?
   \end{itemize}
    \end{itemize}
    \item \textbf{Strategies for understanding familiar and unfamiliar UI}
    \begin{itemize}
           \item How do you go about navigating and interacting with applications or websites?
    \begin{itemize}
     \item What strategy do you use to understand user interfaces?
     \item How do you make sense of UIs?
   \end{itemize}
   \item When you come across an unfamiliar or new website, what do you do to learn it?
    \begin{itemize}
        \item What strategy do you use to understand unfamiliar user interfaces?
        \begin{itemize}
            \item Do you jump to the first heading? Other landmarks?
            \item Do you have other entry strategies?
        \end{itemize}
        \item How do you know where you are in an unfamiliar user interface?
    \end{itemize}
    \end{itemize}
    \item \textbf{Mental models of UI and applications}
    \begin{itemize}
        \item What is your mental model of your operating system?
        \item What mental model do you have for applications (i.e., websites and apps)?    
        \item How important is consistency between different websites or apps to your overall experience?
        \begin{itemize}
            \item Do you have a mental model for specific websites, or broader expected categories? (\eg{} E-commerce for Amazon, Target; Social Media for Twitter, Reddit)
        \end{itemize}
            \item What are important aspects of user interfaces you rely on?
    \end{itemize}
    \item \textbf{Formation and breakdown of expectations in dynamic or inconsistent interfaces}
    \begin{itemize}
        \item When a user interface updates or frequently changes its interface, how do you notice the change?
    \begin{itemize}
        \item What do you do when there’s a mismatch?
        \item How do you adapt to it?
        \item How long might it take to remember the change?
    \end{itemize}
    \item What UI problems do you encounter often?
    \item What happens when a UI breaks your expectations?
        \begin{itemize}
            \item How do you feel about it?
            \item What do you do?
        \end{itemize}
    \end{itemize}
    \item \textbf{Support systems for understanding UI (\eg{} assistance from others or use of AI tools)}
    \begin{itemize}
        \item Do you rely on a sighted individual to understand user interfaces?
        \begin{itemize}
            \item If so, how often? Any particular cases?
        \end{itemize}
    \item Do you communicate often about UI? At work? With family members?
    \item Do you rely on external systems (i.e., AI) to understand user interfaces?
        \begin{itemize}
            \item If so, how often? Any particular cases?
        \end{itemize}
    \end{itemize}
\end{enumerate}

\subsection{Website Walkthrough}
\begin{enumerate}
    \item What is your mental model of this website?
    \begin{itemize}
        \item What is it for? What can you do? What works? What doesn’t work?
    \end{itemize}
    \item In one sentence, could you describe this website?
    \item Could you walk me through this website?
    \item How did you come to understand this website?
    \item What makes sense to you?
    \item What doesn’t make sense to you?
    \item What is missing?
    \item What do you wish you knew? What do you wish was different?
    \item What’s the most important piece of information that you could use?
\end{enumerate}

\section{User Evaluation Questions}
\label{appendix:evaluation}
\subsection{Likert-scale Questions}
\begin{enumerate}
    \item I feel confident describing this website
    \item I understand the overall structure of this website
    \item I understand the functionality of this website
    \item I understand what the website might look like visually
    \item I felt comfortable navigating this website
    \item The behavior of the website matched my expectations
    \item I was able to understand changes without manually re-exploring
\end{enumerate}
\subsection{Reflective Interview Questions}
\begin{enumerate}
    \item First of all, what are your overall impressions of Coral?
    \begin{itemize}
        \item What did you like about Coral? What moments were most helpful for you?
        \item What could have been better about Coral?
    \end{itemize}
        \item How, if at all, did Coral influence your understanding of the interface structure?
    \begin{itemize}
        \item What information was helpful?
        \item Was anything unhelpful?
    \end{itemize}
        \item When you moved between pages, did Coral change how you carried over your understanding from one interface to another?
        \begin{itemize}
            \item What carried over?
            \item What didn’t?
        \end{itemize}
        \item Did Coral ever refer back to something you did earlier (e.g., a previous step, page, or site)?
    \begin{itemize}
        \item If so, was that helpful or confusing?
        \item If not, did you wish it did?
    \end{itemize}
    \item Were there moments when Coral’s descriptions did not match what you expected?
\begin{itemize}
    \item Was that due to Coral or the website?
    \item How did you resolve it?
\end{itemize}
\item Did Coral help you recover when something unexpected happened?
\begin{itemize}
    \item Can you describe a specific example?
    \begin{itemize}
        \item Can you describe what you did before Coral intervened?
        \item What did Coral change in that moment?
    \end{itemize}
\end{itemize}
\item Did Coral reduce the need to manually re-explore the interface?
\begin{itemize}
    \item How?
\end{itemize}
\item Did Coral reduce or increase your mental effort?
\begin{itemize}
    \item Did it ever feel like too much information?
\end{itemize}
\item  How much did you trust Coral’s descriptions?
\begin{itemize}
    \item Were there moments where you explored to double-check?
\end{itemize}
\item Were there moments when Coral spoke too often or not enough?
\item Compared to only using your screen reader, what was the biggest difference for you?
\item Would you want Coral like this to be a part of your screen reader?
\begin{itemize}
    \item In what situations would you want it? Are there any situations where you don’t want it?
    \item If not, are there traits you would like to have in your screen reader?
\end{itemize}
\item If you encountered a similar site tomorrow without Coral, what do you think you would already understand?
\end{enumerate}

\section{Coral Prompts}
\label{appendix:prompts}
\subsection{URL Layout}
\label{appendix:prompt-url}
\begin{lstlisting}
ROLE: Navigation structure describer for screen-reader users.
TARGET URL: ${url}
TASK: Provide one sentence describing likely navigation structure and primary page regions.

RULES:
- Output exactly one sentence.
- Use plain words and avoid developer terms.
- Balance structure and function (where regions are, and what users can do there).
- Skip branding and visual style details.
\end{lstlisting}
\subsection{Screenshot Summary}
\label{appendix:visual}
\begin{lstlisting}
ROLE: Screen layout describer for screen-reader users.
TASK: give a short plain-language summary of page structure and interactive areas.

RULES:
- Max 2 sentences.
- Plain text only.
- Focus on structure first (main area, side panel, top bar, form region, results region).
- Include key interactive zones (buttons, fields, menus, lists) in simple words.
- Do not focus on visual styling or decoration.
\end{lstlisting}
\subsection{Semantic Category \& Keywords}
\label{appendix:prompt-semantic}
\begin{lstlisting}
ROLE: Coral Semantic Architect.
INPUTS: Page metadata, headings, and landmarks.
TASK: Generate a machine-readable structural fingerprint (JSON).

KEYWORD RULES:
- Quantity: Exactly 5 functional keywords.
- Format: hyphenated-kebab-case.
- Focus: "Structural Skeleton" only (Layout, Navigation, Interaction patterns).
- Exclude: Visual styling, branding, or page-specific niche terms.
- Preference: Favor reusable UI patterns (e.g., 'sticky-header', 'grid-layout', 'collapsible-sidebar').

OUTPUT FORMAT (JSON ONLY):
{
  "category": "string",
  "keywords": ["keyword-1", "keyword-2", "keyword-3", "keyword-4", "keyword-5"]
}
\end{lstlisting}
\subsection{Importance Evaluation}
\label{appendix:importance}
\begin{lstlisting}
ROLE: Importance filter for a screen-reader assistant.
GOAL: announce only meaningful changes. Default to false unless clearly useful.

INPUTS:
1. Screen summary (may be reused from previous state).
2. Last user action.
3. Current page context.
4. Change signals (text and node changes).
5. Recent interaction history (may be empty).

SET isImportant=true WHEN:
- Page load or page switch introduces a new task context.
- A change affects what the user can do next (submit, continue, fix an error, navigate).
- The change is the direct result of the last user action.
- A blocking or urgent message appears.

SET isImportant=false WHEN:
- Cosmetic or repetitive updates.
- Passive background updates that do not change user decisions.
- Small text churn with no change in actionability.
- Node-only removals appear without recent interaction intent.

OUTPUT FORMAT (JSON ONLY):
{ "isImportant": boolean }
\end{lstlisting}
\subsection{Gist Synthesis}
\label{appendix:gist}
\begin{lstlisting}
ROLE: Accessibility narrator for blind and low-vision users.

INPUTS:
- LAYOUT CONTEXT: page regions and where changes happened.
- FUNCTION CONTEXT: user trigger and what users can do next.
- CHANGE SIGNALS: explicit text/node deltas when present.
- PREVIOUS STATE: previous URL/screen summary/output (may be null).
- MEMORY CONTEXT: prior similar UI outcomes (may be empty).
- TRANSITION CONTEXT: functional area visibility states and page-to-page toggles.

INPUT FORMAT:
- The user message is compact JSON.
- Treat "layout" and "functionality" as highest-priority evidence.

TASK:
Generate ONE clear sentence that tells the user what is new and where it happened.

MEMORY USAGE:
- If memory is provided and matches current intent, use it to disambiguate what likely changed.
- Never override explicit current-page evidence with memory.
- Use memory as tie-breaker context, not as primary evidence.

CHANGE MODE GUIDANCE:
- If this is a new page or major shift:
  Explain the page in simple spatial terms.
- If this is an incremental update:
  Report only the new information; skip unchanged header/nav boilerplate.
- Landmarks include visibility state:
  treat landmarks with state.visible=false as hidden/unavailable, not active page areas.
- Transition includes functionalAreas and topNavigation:
  rely on transition.topNavigation.visible to decide if a navbar is available.
- If transition.topNavigation.presentInDom=true and transition.topNavigation.visible=false:
  do NOT describe a visible top navigation bar.
- If a functional area (nav, header, sidebar, dialog, menu, toolbar) toggles visibility:
  describe it as shown/hidden only when that change affects navigation or available actions.
- If explicit text/node deltas are present:
  Prioritize them as concrete evidence of change.
- For node changes:
  prioritize functional UI chrome changes first (navigation, sidebar, menu, toolbar, dialog, header).
- Mention removals only when they are functional and change what the user can do.
- If both additions and removals are present:
  lead with functional removals/additions, then briefly mention page content.
- For page-to-page transitions:
  focus on what the current page now offers, not a before/after comparison.
- If transition.pageChanged is true:
  use transition.appearedAreas and transition.visibilityToggles as primary evidence for functional structure changes.
- If region tags are provided (header, nav, main, section):
  use those anchors, and narrate the section tag as "content area".

LANGUAGE RULES:
- Avoid technical UI jargon (avoid words like "modal", "component tree", "DOM", "delta").
- Use plain screen-reader-friendly terms: page, area, list, button, field, message, menu.
- Prefer direct verbs: opened, closed, expanded, collapsed, selected, added, removed, updated.
- Balance structure and interactivity:
  mention both where it is (location/area) and what it does or what changed.
- Prefer current-state phrasing over before/after contrast.
- Avoid repetitive opener templates, especially "The main area now ...".
- Avoid rigid templates; vary sentence openings naturally.

OUTPUT RULES:
- Exactly one sentence.
- No filler like "The state is now" or "You are on".
- Include location when useful, but do not force the sentence to start with a location phrase.
- If additions are present, include an explicit "added" or "now shows" clause.
- Mention removals only when they are functional and meaningful to workflow.
- Do not use strong contrastive phrasing like "no longer shows X and now shows Y" unless X was a critical functional area.
- Keep page content summary brief after functional changes.
\end{lstlisting}